\documentclass[a4paper,twocolumns,11pt]{quantumarticle}
\pdfoutput=1

\usepackage{graphicx}% Include figure files
\usepackage{bm}% bold math
\usepackage{amsmath,amsfonts,amssymb}
\usepackage{physics}
\usepackage{hyperref}
\usepackage{url}
\usepackage{txfonts}
\usepackage{dsfont}
\usepackage{mathtools}

\usepackage[final]{changes}
\usepackage{natbib}

\usetikzlibrary{external}
\begin{document}
\title{Gaussian quantum metrology for mode-encoded parameters}
\author{Giacomo Sorelli} 
\affiliation{Laboratoire Kastler Brossel, Sorbonne Universit\'e, ENS-Universit\'e  PSL, CNRS, Collège de France, 4  Place Jussieu, F-75252 Paris, France}
\affiliation{Fraunhofer IOSB, Ettlingen, Fraunhofer Institute of Optronics,
System Technologies and Image Exploitation, Gutleuthausstr. 1, 76275 Ettlingen, Germany} 
\author{Manuel Gessner}
\affiliation{Departament de F\'isica Te\`orica, IFIC, Universitat de Val\`encia, CSIC, C/ Dr. Moliner 50, 46100 Burjassot (Val\`encia), Spain}
\author{Nicolas Treps}
\affiliation{Laboratoire Kastler Brossel, Sorbonne Universit\'e, ENS-Universit\'e PSL, CNRS, Collège de France, 4  Place Jussieu, F-75252 Paris, France}
\author{Mattia Walschaers} 
\affiliation{Laboratoire Kastler Brossel, Sorbonne Universit\'e, ENS-Universit\'e PSL, CNRS, Collège de France, 4  Place Jussieu, F-75252 Paris, France}

%\date{\today}
\begin{abstract}
Quantum optical metrology aims to identify ultimate sensitivity bounds for the estimation of parameters encoded into quantum states of the electromagnetic field. 
In many practical applications, including imaging, microscopy, and remote sensing, the parameter of interest is not only encoded in the quantum state of the field, but also in its spatio-temporal distribution, i.e. in its mode structure.
In this mode-encoded parameter estimation setting, we derive an analytical expression for the quantum Fisher information valid for arbitrary multimode Gaussian fields.
To illustrate the power of our approach, we apply our results to the estimation of the transverse displacement of a beam and to the temporal separation between two pulses. 
For these examples, we show how the estimation sensitivity can be enhanced by adding squeezing into specific modes. 
\end{abstract}
\maketitle

\section{Introduction}
A fundamental task in quantum metrology is to identify the ultimate sensitivity limit in the estimation of a parameter encoded into a quantum state.
Even under ideal conditions, when all technical noise sources are removed, quantum noise poses unavoidable limitations to such estimation.
In spite of that, quantum parameter estimation theory provides the tools to reduce noise by optimizing the output measurements.
This optimization leads to the quantum Cram\'er Rao lower bound, which states that the minimal uncertainty $\Delta \vartheta$ of the estimator of a parameter $\vartheta$ is inversely proportional to the quantum Fisher information of the quantum state $\hat{\rho}_\vartheta$ where the parameter is encoded
\citep{helstrom1976,holevo2011probabilistic,Paris2009, Luca_Augusto_review,GiovannettiLoydMaccone,Toth_2014}.
This bound can be further optimized by finding quantum states that, for a given parameter, maximize the value of the quantum Fisher information.

Electromagnetic fields play a privileged role as metrological probes in a variety of branches of science and technology, ranging from imaging and microscopy \citep{TAYLOR20161,Tsang2019}, to remote sensing with lidars and radars \citep{Giovannetti2001,Zhuang_Lidar,HuangPRXQuantum}, to gravitational wave detection \citep{PhysRevLett.123.231108,PhysRevLett.123.231107}.
In several of these applications, the parameter of interest does not only modify the quantum state of the probe light, but also its spatio-temporal distribution.
Such a spatio-temporal distribution is conveniently described in terms of {\it modes}. i.e. normalized solutions of Maxwell's equations in vacuum \citep{Fabre:2020}.
For example, spatial modes of light describe the different components of an image, while the properties of an optical pulse are encoded into frequency-time modes.

Previous works in this context of {\it mode-encoded parameter estimation} focused on specific problems.
For example, the case where the total light's intensity is not affected by the parameter, but its distribution among different modes is, was considered for the estimation of a small lateral beam displacement \citep{TrepsPRL2002,TrepsScience2003}, or in the estimation of spectral parameters of a frequency comb \citep{CaiNPJquantum2021}.
A general theory for this fixed-intensity scenario was recently presented by \cite{gessner2022quantum}.
Two different (mathematically equivalent) problems, that lately attracted a lot of attention, are the estimation of the separation between two point sources analysed through a diffraction-limited imaging system \citep{Tsang_PRX,Paur:16,Boucher:20} or the temporal separation between two pulses \citep{Ansari_PRXQuantum_2021,De_PRRes_2021,Mazelanik_nat_comm_22}.
For these problems, the parameter of interest is encoded in the shape of two (spatial or temporal) modes in the detection plane with separation-dependent populations \citep{LupoPirandola}.
Pushed by the need to go beyond these case studies, in this work, we study mode-encoded parameter estimation with arbitrary multimode Gaussian states, i.e. photonics quantum states fully defined by the first two moments of their quadratures \citep{Holevo:1975,Weedbrook:2012,Adesso:2014}.

Gaussian states play a central role in quantum optics: they describe important classical states such as coherent states, representing lasers operating above threshold, and thermal states, describing fully incoherent light. 
Furthermore, non-classical Gaussian states can be produced deterministically in non-linear optical processes.
Among the latter states, there are squeezed states, whose reduced quantum noise has been proposed as a useful resource since the early days of quantum parameter estimation \citep{Caves1981}, and is now a key ingredient of several quantum-enhanced metrological schemes \citep{TrepsScience2003,
PezzePRL2008,PhysRevLett.123.231108,PhysRevLett.123.231107}. 
While previous studies of the quantum Fisher information for Gaussian states exist \citep{PinelPRA2012,monras2013, vsafranek2015,JiangPRA2014},
they focused on the estimation of parameters defining the first two moments of the quadratures, e.g. mean field, phase, and squeezing.

The aim of this work is to overcome these limitations, to study the estimation of parameters encoded in the spatio-temporal profile of the electromagnetic field, and therefore to broaden the applicability of Gaussian quantum metrology to new fields of technology such as imaging, microscopy, and temporal (or spectral) beam profiling.
As examples of such applications, we reconsider the estimation of the transverse displacement of a beam and the temporal separations between two pulses:
For the former case, we extend the results for coherent beams of \citep{PinelPRA2012} to thermal beams, we show that squeezing in the right mode provides a quantum enhancement also in this case, and we discuss how to include the effect of thermal noise and losses.
For the latter, we confirm known results for thermal \citep{Nair_2016, LupoPirandola} and coherent pulses \citep{Sorelli_PRRes_2022}, and we investigate the possibility of a quantum enhancement populating additional modes with squeezed light. 

Our paper is organized as follows: First, in Sec.~\ref{Sec:Preliminaries}, we recall some basic facts about Gaussian states and quantum parameter estimation. 
We then derive the analytical expression of the quantum Fisher information for mode-encoded parameter estimation with Gaussian states, in Sec.~\ref{Sec:mode_enc_param_est}.
Section~\ref{sec:examples} contains the application of our results to the estimation of the transverse displacement of a beam and the temporal separation of two pulses. Section~\ref{Sec:conclusion} concludes our work.

\section{Preliminaries}
\label{Sec:Preliminaries}
\subsection{Gaussian states}
An $N-$mode continuous variable (CV) \citep{Braunstein:2005,serafini_book}  quantum system can be described by choosing a {\it mode basis}, i.e. a set $\left\{ u_k({\bf r}, t) \right\}_{k=1}^N$ of solutions of Maxwell's equations, orthonormal with respect to the inner product 
\begin{equation}
\left(u_k|u_l\right) = \int d^3{\bf r} u_k^*({\bf r}, t)  u_l({\bf r}, t)  = \delta_{kl},
\end{equation}
and associating with each mode $u_k ({\bf r}, t)$ a pair of quadrature operators $\hat{q}_k = (\hat{a}_k +\hat{a}_k^\dagger)$ and $\hat{p}_k = i(\hat{a}_k^\dagger - \hat{a}_k)$, where $\hat{a}_k^\dagger$ and $\hat{a}_k$ are standard creation and annihilation operators.
If we group all quadratures in the $2N-$dimensional vector $\hat{\bf x} = (\hat{q}_1, \hat{p}_1, \dots, \hat{q}_n, \hat{p}_n)^\top$, from the canonical commutation relations $[\hat{a}_k,\hat{a}^\dagger_l] = \delta_{kl}$ for annihilation operators, we obtain  
\begin{equation}
[\hat{x}_j,  \hat{x}_k] = 2i \Omega_{jk}, 
\end{equation}
with the symplectic form $\Omega = \bigoplus_{k=1}^N \omega_k$, and $\omega_k = i \sigma_y$, where we have introduced the notation $\sigma_{i=x,y,z}$ for the standard $2\times 2$ Pauli matrices. 
Our preferred {\it phase space} representation of an $N-$mode CV quantum state with density matrix $\hat{\rho}$, and  characteristic function $\chi({\bf y}) = \tr[\exp\left(-i {\bf y}^\top \Omega \hat{\bf x}\right)\hat{\rho}]$, is the Wigner function 
\begin{equation}
W({\bf x}) = \int \frac{d {\bf y}^{2N}}{(2\pi)^{2N}} e^{-i {\bf y}^\top\Omega {\bf x}}\chi({\bf y}).
\end{equation} 

In this work, we restrict ourselves to the study of Gaussian states.
$N-$mode Gaussian states are CV states with Gaussian Wigner function \citep{Holevo:1975,Weedbrook:2012,Adesso:2014}
\begin{equation}
W ({\bf x}) = \frac{\exp \left[- ({\bf x} -  \bar{\bf x} )^\top\sigma^{-1}({\bf x} -  \bar{\bf x} )/2\right]}{(2\pi)^N \sqrt{\det \sigma}},
\label{Gaussian_W}
\end{equation}
which are completely determined by the displacement vector $\bar{\bf x} = \langle \hat{\bf x} \rangle$, and the covariance matrix
\begin{equation}
{\bf \sigma}_{jk}  = \frac{1}{2}\left\langle\left\{(\hat{x}_j -\bar{ x}_j), (\hat{x}_k -\bar{x}_k)\right\} \right\rangle,
\end{equation}
where $\{\cdot, \cdot\}$ denotes the anticommutator, and $\langle \cdot \rangle$ the expectation value $\langle \cdot \rangle = \tr \left[  \cdot  \hat{\rho} \right]$. 
For every physical state, the covariance matrix satisfies the uncertainty inequality \citep{Simon:1994}
\begin{equation}
{\bf \sigma} + i\Omega \geq 0.
\label{uncertainty}
\end{equation}
A useful property of the covariance matrix is that, according to Williamson theorem \citep{Williamson:1936}, it can be decomposed as 
\begin{equation}
{\bf \sigma} = S {\bf \nu} S^\top, \quad {\rm with} \quad {\bf \nu} = \bigoplus_{k=1}^{N} \nu_k \mathds{1}_2,
\label{Williamson}
\end{equation}
where $S$ is a symplectic matrix, i.e. $S\Omega S^\top = \Omega$, we introduced the notation $\mathds{1}_n$ for the $n-$dimensional identity matrix, and ${\bf \nu}$  is a diagonal matrix whose element are known as {\it symplectic eigenvalues}.
The uncertainty inequality~\eqref{uncertainty} implies that the symplectic eigenvalues must be larger than unity, i.e. $\nu_k \geq 1$. 

\subsection{Mode and state transformations}
\label{Sec:modes_states}
Let us recall that the space of solutions of Maxwell's equation is a Hilbert space \citep{Fabre:2020,Walschaers:2021}. 
Accordingly, different mode bases are connected via unitary transformations
\begin{subequations}
\begin{align}
u_k({\bf r},t) &= \sum_{l=1}^N U_{kl} v_l({\bf r},t),\\
v_k({\bf r},t) &= \sum_{l=1}^N U_{lk}^* u_l({\bf r},t),
\end{align}
\end{subequations} 
with $U_{kl} = \left( v_l | u_k \right)$. 
Under mode basis changes, creation operators follow the same transformation rules as the modes \citep{Fabre:2020}, which implies that the quadrature vector transforms according to 
\begin{equation}
\hat{\bf x}^\prime = O \hat{\bf x},
\label{mode_change}
\end{equation}
with $O$ an orthogonal symplectic matrix, i.e $OO^\top  =O^\top O =  \mathds{1}_{2N}$ and $O^\top\Omega O = \Omega$, with elements $O_{2k-1,2l-1} = \Re\left[U_{kl}\right]$, $O_{2k,2l-1} = -\Im \left[U_{kl}\right]$, $O_{2k-1,2l} = \Im \left[U_{kl}\right]$ and $O_{2k,2l} = \Re\left[U_{kl}\right]$.

A mode basis change is a particular case of a Gaussian channel: a completely positive, trace preserving map transforming Gaussian states into Gaussian states. Such channels are completely determined by their transformation rules for the displacement vector and the covariance matrix \citep{Holevo:2001}
\begin{subequations}
\begin{align}
 \bar{\bf x}^\prime &= \mathcal{T} \bar{\bf x} + \bar{\bf z},\\ 
 {\bf \sigma}^\prime &= \mathcal{T} {\bf \sigma} \mathcal{T}^\top + \mathcal{N},
\end{align}
\label{Gaussian_channel}
\end{subequations}
where $\bar{\bf z}$ is a real $2N-$dimensional vector, while $\mathcal{T}$ and $\mathcal{N}$ are $2N \times 2N$ real matrices with $\mathcal{N} = \mathcal{N}^\top$ and satisfying the positivity condition $\mathcal{N} +i\mathcal{T}\Omega \mathcal{T}^\top \geq i\Omega$.
From Eq.~\eqref{mode_change} is easy to see that a mode basis change is a Gaussian channel (see Eqs.~\eqref{Gaussian_channel}) with $\mathcal{N} =0$ and $\mathcal{T}=O$.

\subsection{Quantum estimation theory}
\label{sec:Q_est_theory}
Let us now assume that we want to estimate a parameter $\vartheta$ encoded in a quantum state $\hat{\rho}_\vartheta$ from $M$ independent measurements of a given positive operator-valued measure (POVM) defined by the operators $ \hat{K}_\mu$. 
Using classical post-processing techniques, from the measurements' results, we can extract an unbiased estimator $\tilde{\vartheta}$ of the parameter as well as its standard deviation $\Delta \tilde{\vartheta}$.
The latter is bounded according to the Cram\'er-Rao inequality \citep{helstrom1976, holevo2011probabilistic,  Paris2009, Luca_Augusto_review, GiovannettiLoydMaccone} 
\begin{equation}
\Delta \tilde{\vartheta} \geq \frac{1}{\sqrt{M \mathcal{F}_{\vartheta,\hat{K}_\mu }}}
\end{equation}
with the Fisher information $\mathcal{F}_{\vartheta,\hat{K}_\mu }$ defined by
\begin{equation}
\mathcal{F}_{\vartheta,\hat{K}_\mu} = \sum_{\mu} p(\mu|\vartheta)\left[\partial_\vartheta \log\left( p(\mu|\vartheta) \right) \right]^2
\end{equation}
where $p(\mu|\vartheta) = \Tr \left[ \hat{K}_\mu \hat{\rho}_\vartheta \right]$ is the conditional probability of obtaining the result $\mu$ for a given value of $\vartheta$, and we introduced the compact notation $\partial_\vartheta \cdot = \partial \cdot / \partial \vartheta$ for the derivative.
The Fisher information optimized over all possible POVMs
\begin{equation}
F_{\vartheta} = \max_{\hat{K}_\mu} \mathcal{F}_{\vartheta,\hat{K}_\mu},
\end{equation}
is the quantum Fisher information (QFI), and establishes the ultimate metrological sensitivity \citep{BraunsteinCaves1994}.
In general, the QFI can be computed as 
\begin{equation}
F_{\vartheta} = \Tr \left[\hat{\mathcal{L}}^2_\vartheta \hat{\rho}_\vartheta\right],
\label{QFI}
\end{equation}
where $\hat{\mathcal{L}}_\vartheta$ is the symmetric logarithmic derivative (SLD), implicitly defined by the equation \citep{helstrom1976, holevo2011probabilistic, Paris2009, Luca_Augusto_review, GiovannettiLoydMaccone} 
\begin{equation}
2\partial_\vartheta \hat{\rho}_\vartheta = \hat{\mathcal{L}}_\vartheta \hat{\rho}_\vartheta + \hat{\rho}_\vartheta \hat{\mathcal{L}}_\vartheta.
\end{equation} 

When $\rho_\vartheta$ is an $N-$mode Gaussian state defined by the displacement vector $\bar{\bf x}$ and the covariance matrix $\sigma$, the SLD is quadratic in the quadratures \citep{monras2013, vsafranek2015}
\begin{equation}
\hat{\mathcal{L}}_\vartheta = L^{(0)}_\vartheta + {\bf L}^{(1)\top}_\vartheta \hat{\bf x} + \frac{1}{2}\hat{\bf x}^\top L^{(2)}_\vartheta \hat{\bf x}
\label{SLD}
\end{equation}
with
\begin{subequations}
\begin{align}
 L^{(0)}_\vartheta &= -\frac{1}{2} \Tr \left[\sigma L_\vartheta^{(2)} \right] - {\bf L}^{(1)\top}_\vartheta \bar{\bf x} - \frac{1}{2}\bar{\bf x}^\top L^{(2)}_\vartheta \bar{\bf x} \\ 
{\bf L}^{(1)}_\vartheta  &= \sigma^{-1}\left(\partial_\vartheta \bar{\bf x} \right) -  L^{(2)}_\vartheta \bar{\bf x}\\
L^{(2)}_\vartheta &=\frac{1}{2}\sum_{l=0}^3\sum_{jk=1}^N \frac{a_{jk}^{(l)}}{\nu_j\nu_k - (-1)^l} \left(S^\top\right)^{-1} A_{jk}^{(l)} S^{-1} \label{L2}
\end{align}
\label{SLD_terms}
\end{subequations}
where $S$ and $\nu$ are the symplectic matrix and the symplectic eigenvalues obtained from the Williamson decomposition of $\sigma$ as introduced in Eq.~\eqref{Williamson}, while $a_{jk}^{(l)} = \Tr \left[ A_{jk}^{(l)}  S^{-1}  (\partial_\vartheta \sigma) (S^\top)^{-1}\right]$, with $A_{jk}^{(l)}$ a set $2n \times 2n$ matrices that are zero everywhere except in the $jk$ block where they are given by $\omega/\sqrt{2}$, $\sigma_z/\sqrt{2}$, $\mathds{1}_2/\sqrt{2}$ and $\sigma_x/\sqrt{2}$ for $l=0,1,2$ and $3$, respectively.
Substituting Eqs.~\eqref{SLD} and \eqref{SLD_terms} into Eq.~\eqref{QFI} and using the properties of the characteristic function \citep{serafini_book}, we can write the QFI for a Gaussian state as 
\begin{equation}
F_\vartheta = F_\sigma + F_{\bar{\bf x}},
\label{QFI_gauss}
\end{equation}
with 
\begin{subequations}
\begin{align}
F_\sigma &= \Tr \left[ L^{(2)}_\vartheta (\partial_\vartheta \sigma)\right] \label{trL2}  \\
&= \frac{1}{2}\sum_{l=0}^3\sum_{jk=1}^N \frac{\left(a_{jk}^{(l)}\right)^2}{\nu_j\nu_k - (-1)^l}, \nonumber \\
F_ {\bar{\bf x}} &= (\partial_\vartheta \bar{\bf x})^\top \sigma^{-1} (\partial_\vartheta \bar{\bf x}). \label{F_x}
\end{align}
\end{subequations}
where $F_\sigma$ and $F_{\bar{\bf x}}$ are the contribution to the QFI coming from variations of the covariance matrix $\sigma$ and the displacement vector $\bar{\bf x}$, respectively.

\section{Mode-encoded parameter estimation}
\label{Sec:mode_enc_param_est}
Let us consider the estimation of a parameter $\vartheta$ encoded into a Gaussian state $\hat{\rho}_\vartheta$ expressed into an $n-$dimensional mode basis $u_k[\vartheta]({\bf r}, t)$, with $n$ the smallest number of modes necessary to describe the system. 
We will refer to the Hilbert space spanned by these modes as $\mathcal{H}_n = {\rm span} \left( \{u_k[\vartheta]({\bf r}, t) \}\right)$.
Since every basis of the mode Hilbert space $\mathcal{H}_n$ would provide a description of the quantum state  $\hat{\rho}_\vartheta$ in terms of the smallest number $n$ of modes, the choice of the mode basis $u_k[\vartheta]({\bf r}, t)$  is not unique. 
Despite this freedom of choice, in the most general parameter estimation scenarios, every mode basis $u_k[\vartheta]({\bf r}, t)$ will be parameter-dependent.
The latter fact implies that the Gaussian state $\hat{\rho}_\vartheta$ depends on $\vartheta$ not only explicitly through the displacement vector $\bar{\bf x}_\vartheta$ and the covariance $\sigma_\vartheta$, but also implicitly through the mode functions $u_k[\vartheta]({\bf r}, t)$.
Our goal is this section is to calculate the QFI~\eqref{QFI_gauss} taking into account both these dependences.

\subsection{Separation of mode and state parameter dependence}
Our first step is to make the parameter dependence coming from the modes $u_k[\vartheta]({\bf r}, t)$ explicit in the covariance matrix and displacement vector of the quantum state $\hat{\rho}_\vartheta$. 
To this goal, we express them into a parameter-independent basis $v_k({\bf r}, t)$: Using Eq.~\eqref{mode_change}, we get
\begin{subequations}
\begin{align}
\sigma_I &= O \sigma_\vartheta O^T \label{mode_independent_sigma},\\
\bar{\bf x}_I &= O \bar{\bf x}_\vartheta, \label{mode_independent_x}
\end{align}
\label{mode_independent}
\end{subequations}
where we introduced the subscripts $I$ and $\vartheta$ to identify quantities in the parameter-independent and parameter-dependent bases, respectively.
Naturally, the choice of the parameter-independent basis $v_k({\bf r}, t)$ is not unique. However, since this basis does not contain any information on the parameter, its choice does not affect the final expression for the QFI, as will become clear at the end of our calculation.

Given that $n$ is the smallest number of modes necessary to represent the state $\hat{\rho}_\vartheta$, the parameter independent basis $v_k({\bf r}, t)$ must have dimension $N \geq n$.
To take into account this change in dimension, we complement the state in the parameter-dependent mode basis with $N-n$ vacuum modes, so that we can write the covariance matrix $\sigma_\vartheta$ in block diagonal form as
\begin{equation}
\sigma_\vartheta = 
\begin{pmatrix}
V_n & 0 \\
0  & \mathds{1}_{2(N-n)}
\end{pmatrix},
\label{sigma_V}
\end{equation}
and the displacement vector as $\bar{\bf x}_\vartheta = (\bar{\bf x}^\top_n, 0, \cdots, 0)^\top$.
To isolate the action of $O$ on the $n$ initially populated modes, it is convenient to rewrite it as a $1\times 2$ block matrix 
\begin{equation}
O = 
\begin{pmatrix}
O_n &
O_{N-n}
\end{pmatrix},
\label{O_block}
\end{equation}
with $O_n$ and $O_{N-n}$ a $2N \times 2n$ and a $2 N \times  2(N-n)$ matrices, respectively. 
Some useful properties of these matrices and their derivatives are reported in  App.~\ref{App:OnON}.
Substituting Eqs.~\eqref{O_block} and \eqref{sigma_V} into Eq.~\eqref{mode_independent_sigma}, and using the properties of the matrices $O_n$ and $O_{N-n}$ (See Eq.~\eqref{O_properties_4} in  App.~\ref{App:OnON}), we can rewrite the covariance matrix in the mode-independent basis as
\begin{equation}
\sigma_I = O_n(V_n - \mathds{1}_{2n})O^\top_n +\mathds{1}_{2N}.
\label{V_On}
\end{equation}
Analogously, using Eq.~\eqref{O_block} into Eq.~\eqref{mode_independent_x}, we can rewrite the displacement vector as 
\begin{equation}
\bar{\bf x}_I = O_n \bar{\bf x}_n.
\label{x_On}
\end{equation}

Equations~\eqref{V_On} and \eqref{x_On} provide a description of the Gaussian state $\hat{\rho}_\vartheta$ where the parameter dependence is fully expressed in the covariance matrix $\sigma_I$ and the displacement vector ${\bf x}_I$.
In particular, the transformation properties of the $n$ initially populated modes appear now explicitly through the matrix $O_n$.
In the following, we are going to use these expressions to compute the two terms in Eq.~\eqref{QFI_gauss}.

\subsection{Covariance matrix contribution to the quantum Fisher information}
\label{sec:covariance_mat_contr}
We start with the calculation of $F_\sigma$ (see Eq.~\eqref{trL2}), which describes the contribution to the sensitivity due to variations of the covariance matrix.
Let us start by taking the derivative of the covariance matrix $\sigma_I$ in the parameter independent basis with respect to the parameter 
\begin{align}
\partial_\vartheta \sigma_I &= \left(\partial_\vartheta O_n \right)(V_n - \mathds{1}_{2n})O^\top_n \\
&\; + O_n (V_n - \mathds{1}_{2n})\left(\partial_\vartheta O^\top_n\right) +O_n  \left(\partial_\vartheta V_n\right) O^\top_n \nonumber.
\end{align}
To compute the quadratic term of the SLD $L^{(2)}_\vartheta$ (see Eq.~\eqref{L2}), we need the Williamson decomposition $\sigma_I = S_I \nu_I S^T_I$ of the covariance matrix $\sigma_I$. 
Using Eqs.~\eqref{sigma_V} and \eqref{O_block}, we can connect it to the Williamson decomposition $V_n = S_n \nu S^T_n$ of the covariance of the $n$ initially populated modes in the parameter dependent basis $u_n[\vartheta]({\bf r},t)$, and obtain
\begin{subequations}
\begin{align}
S_I &= \begin{pmatrix}
O_n S_n & O_{N-n}
\end{pmatrix}, 
\label{S}\\ 
\nu_I &= \nu \oplus \mathds{1}_{2(N-n)}.
\label{nu}
\end{align} 
\end{subequations}
Accordingly, using the properties of the matrix $O$ (see  App.~\ref{App:OnON} for details), we can write 
\begin{equation}
S^{-1}_I \left(\partial_\vartheta \sigma_I \right)\left(S^\top_I\right)^{-1} = 
\begin{pmatrix}
B_n & B_{\partial}^\top & 0 \\
B_{\partial} & 0 & 0 \\ 
0 & 0 & 0
\end{pmatrix},
\label{SparVS}
\end{equation}
with 
\begin{subequations}
\begin{align}
B_n =\:& S_n^{-1}D^\top_n\left( V_n - \mathds{1}_{2n}\right) \left(S_n^\top \right)^{-1} \nonumber \\ 
& +S_n^{-1}\left( V_n - \mathds{1}_{2n}\right)D_n   \left(S_n^\top\right)^{-1} \label{Bn} \\
& +S_n^{-1}\left(\partial_\vartheta V_n\right) \left(S_n^\top \right)^{-1}, \nonumber \\
B_{\partial} =\:& D_\partial^\top\left( V_n - \mathds{1}_{2n}\right) \left(S_n^\top\right)^{-1}.\label{Bpartial}
\end{align}
\label{B}
\end{subequations}
Here, $D_{n}$ is a $2n \times 2n$ matrix and $ D_{\partial}$ is a $2n \times 2m$ matrix, constructed using, respectively, the coefficients $c_{kl}[\vartheta]$ and $c^\prime_{kl}[\vartheta]$ of the expansion 
\begin{align}
\partial_\vartheta u_k[\vartheta]({\bf r}, t) &= \sum_{l=1}^n c_{kl}[\vartheta] u_l[\vartheta]({\bf r}, t) \label{derivative_modes}  \\ &\quad+\sum_{l=1}^m c^\prime_{kl}[\vartheta] u^\prime_l[\vartheta]({\bf r}, t),
\nonumber
\end{align}
where the modes $u^\prime_l[\vartheta]({\bf r}, t)$ form an $m(\leq n)-$dimensional basis of the mode Hilbert space $\mathcal{H}_{\partial}={\rm span} \left(\{\partial_\vartheta u_k[\vartheta]({\bf r}, t)\}\right) \setminus \mathcal{H}_n$  (See App.~\ref{App:OnON}).
Accordingly, the diagonal block $B_n$ contains a {\it mode contribution} (first two terms in Eq.~\eqref{Bn}) due to the portion of the derivatives $\partial_\vartheta u_k[\vartheta]({\bf r}, t)$ within the space of the initially populated modes $\mathcal{H}_n$, and a contribution given by the explicit dependence of the covariance matrix $V_n$ on the parameter.
On the other hand, the off-diagonal blocks $B_\partial$ and $B_\partial^\top$ only contain the  mode contribution due to the leakage of the derivatives $\partial_\vartheta u_k[\vartheta]({\bf r}, t)$  from $\mathcal{H}_n$ to $\mathcal{H}_\partial$.

Using Eq.~\eqref{SparVS}, we can calculate the coefficients $a_{jk}^{(l)}$ in Eq.~\eqref{L2}, which result in 
\begin{equation}
a_{jk}^{(l)} = 
\begin{cases}
\Tr \left[A_{jk}^{(l)} B_n \right]  & 1 \leq j,k \leq n\\
\Tr \left[\tilde{A}_{jk}^{(l)\top} B_\partial \right]  & 1 \leq j \leq n, n < k \leq (n+m)\\
\Tr \left[\tilde{A}_{jk}^{(l)} B_\partial^\top \right]  & n < k \leq (n+m), 1 \leq j \leq n \\
0 & j,k > (n+m)
\end{cases}
\label{a}
\end{equation}
where $\tilde{A}_{jk}^{(l)}$ are $m\times n$ blocks of the matrices $A_{ij}^{(l)}$.
Finally, using Eq.~\eqref{trL2} and Eq.~\eqref{nu}, we can write the covariance matrix contribution to the QFI as
\begin{align}
F_\sigma &=\frac{1}{2} \sum_{l=0}^3\sum_{jk=1}^n \frac{\left(a_{jk}^{(l)}\right)^2}{\nu_j\nu_k- (-1)^l} \label{QFI_sigma} \\
&\quad +\frac{1}{2} \sum_{l=0}^3\sum_{j=1}^{n}\sum_{k=1}^{m} \frac{\left(a_{j,k+n}^{(l)}\right)^2 +\left(a_{k+n,j}^{(l)}\right)^2}{\nu_j- (-1)^l}  \nonumber.
\end{align}
The sum in the first term in Eq.~\eqref{QFI_sigma} only runs over the $n$ initially populated modes.
Accordingly, it describes the contribution to the QFI given by variations of the state within the $n$ initially populated modes $u_n[\vartheta]({\bf r},t)$.
On the other hand, the second term in Eq.~\eqref{QFI_sigma} contains a sum over the $n$ initially populated modes $u_n[\vartheta]({\bf r},t)$ and another over their $m$ orthonormalized derivatives $u^\prime_n[\vartheta]({\bf r},t)$. 
Therefore, it takes into account the contribution to the QFI due to the coupling between the initially populated modes and their derivatives induced by parameter variations.
Finally, let us note that Eq.~\eqref{QFI_sigma} is completely determined by the covariance matrix $V_n$ of the state $\hat{\rho}_\vartheta$ in the $n$ initially populated modes $u_k[\vartheta]({\bf r},t)$, and by the shape of the modes themselves, but, as anticipated, it does not depend on the choice of the auxiliary parameter-independent basis $v_k({\bf r}, t)$.

\subsection{Displacement vector contribution to the quantum Fisher information}
We now move on to compute $F_{\bar{\bf x}}$, as given by Eq.~\eqref{F_x}, which takes into account the contribution to the QFI coming from variations of the displacement vector $\bar{x}_I$.
To compute this term, we need the derivative of Eq.~\eqref{x_On}
\begin{equation}
\partial_\vartheta \bar{\bf x}_I = (\partial_\vartheta O_n) \bar{x}_n + O_n(\partial_\vartheta \bar{\bf x}_n),
\label{par_x}
\end{equation}
and the inverse of the covariance matrix $\sigma_I$ that, thanks to Eq.~\eqref{O_block}, we can write as 
\begin{equation}
\sigma_I^{-1} = O_n V_n^{-1}  O^\top_n +  O_{N-n} O^\top_{N-n}.
\label{inverse_sigma}
\end{equation}
Finally, combining Eqs.~\eqref{par_x} and \eqref{inverse_sigma}, and using the properties of the matrices $O_n$ and $O_{N-n}$ (see App.~\ref{App:OnON}), we obtain 
\begin{align}
F_{\bar{\bf x}} &= (\partial_\vartheta \bar{\bf x}_n)^\top V^{-1}_n (\partial_\vartheta \bar{\bf x}_n) \label{Fx} \\
&\;+ (\partial_\vartheta \bar{\bf x}_n)^\top V^{-1}_n D_n^\top \bar{\bf x}_n + \bar{\bf x}^\top_n D_n V^{-1}_n(\partial_\vartheta \bar{\bf x}_n) \nonumber \\ 
&\;+\bar{\bf x}^\top_n\left( D_n V^{-1}_n D^\top_n + D_\partial D^\top_\partial\right)\bar{\bf x}_n. \nonumber
\end{align}
Similarly to what we observed for $F_\sigma$, $F_{\bar{\bf x}}$ only depends on the displacement vector $ \bar{\bf x}_n$ in the $n$ initially populated modes $u_k[\vartheta]({\bf r},t)$ and their shapes.
Moreover, we note that the first term in in Eq.~\eqref{Fx} only depends on variations of the displacement vector $\bar{\bf x}_n$, while the last term only depends on changes of the shapes of the $n$ initially populated modes $u_k[\vartheta]({\bf r}, t)$. 
On the other hand, in the two middle terms appear both $(\partial_\vartheta \bar{\bf x}_n[\vartheta])$ and $D_n$. Accordingly, they combine mode variations with changes in the displacement vector.

\section{Application to spatial and temporal resolution}
\label{sec:examples}
\subsection{Spatial beam positioning}
\label{sec:spatial_positioning}
\subsubsection{A single populated mode}
As a first example, we consider the estimation of the transverse displacement $d$ of a light beam whose spatial profile is defined by the mode $u_0[d]({\bf r}) = u({\bf r} - {\bf r}_0)$ with ${\bf r}_0 = (d ,0)$, where, without loss of generality, we assumed the beam to be displaced along the $x$ axis. 
Furthermore, we consider the mode $u[d]({\bf r})$ to have a well-defined parity, s.t. it is orthogonal to its derivative: $(\partial_d u_0|u_0) = 0$. 
Under these assumptions, we have $D_n = 0$ and $D_\partial = \eta \mathds{1}_2$, with $\eta = ||\partial_d u({\bf r}- {\bf r}_0)||$. In this context, $\eta$ quantifies the spatial extent of the beam we want to localize, e.g. for a Gaussian mode $u({\bf r}) = \exp(-|{\bf r}|^2/2w^2)/\sqrt{\pi w^2}$, we have $\eta^2 = 1/2w^2$.

This estimation problem is fully defined by the mode $u_0[d]({\bf r})$. As a consequence, the mean field contribution to the QFI~\eqref{Fx} simplifies to 
\begin{equation}
F_{\bar{\bf x}} = (\partial_d {\bf x}_0)^\top V_0^{-1}  (\partial_d {\bf x}_0) + \eta^2 ||{\bf x}_0||^2,
\label{F_x_u0}
\end{equation} 
and we can write the covariance contribution to the QFI~\eqref{QFI_sigma} as 
\begin{equation}
F_v = \frac{1}{2} \sum_{l=0}^3  \left( \frac{a_l^2}{\nu_0^2 -(-1)^l} +  \frac{b_l^2}{\nu_0 -(-1)^l}  \right),
\label{F_v_u0}
\end{equation}
where we defined the coefficients
\begin{subequations}
\begin{align}
a_l^2 &= \tr \left[A_l S_0^{-1} (\partial_dV_0)  (S_0^\top)^{-1} \right]^2,  \\
b_l^2 &= \eta^2 \tr \left[A_l  \left( V_0 - \mathds{1}_{2}\right) \left(S_0^\top\right)^{-1} \right]^2 \\
& \quad + \eta^2 \tr \left[A_l S_0^{-1} \left( V_0 - \mathds{1}_{2}\right)  \right]^2, \nonumber
\end{align}
\label{ab}
\end{subequations}
with
\begin{align}
A_0 &= i \sigma_y/\sqrt{2}; \; A_1 = \sigma_z/\sqrt{2}; \\ \nonumber A_2 &= \mathds{1}_2/\sqrt{2}; \;  A_3 = \sigma_x/\sqrt{2},
\end{align}
where we recall $\sigma_{x,y,z}$ are Pauli matrices.  

We can now evaluate Eqs.~\eqref{F_x_u0} and \eqref{F_v_u0} for different states of the mode $u_0[d]({\bf r})$. Let us start by considering a coherent state $\ket{\alpha}$, defined by the complex amplitude $\alpha$ that can be parameter dependent.
Accordingly, we have $\bar{\bf x}_0 = 2(\Re [\alpha], \Im [\alpha])$ and $V_0 = \mathds{1}_2$. 
In this case, is not hard to verify that the covariance matrix contribution~\eqref{F_v_u0} vanishes, $F_v = 0$, and the QFI is fully determined by the displacement term~\eqref{F_x_u0}, which reduces to
\begin{equation}
    F_{d, {\rm coh}} = |\partial_d \alpha|^2 + 4\eta^2 N_0,
    \label{F_x_uo_coherent}
\end{equation}
where we introduced the mean photon number $N_0 = |\alpha|^2$. The second term in Eq.~\eqref{F_x_uo_coherent} presents a shot-noise scaling and is inversely proportional to the beam size: small displacements of a larger beam are harder to estimate. On the other hand, the first term in Eq.~\eqref{F_x_uo_coherent} takes into account how $\alpha$ depends on the transverse displacement of the beam. 
Such a dependence could be induced by position-dependent losses.

Let us now consider the localization of a thermal beam, for which we have $\bar{\bf x}_0 =0$ and $V_0 = (2N_0 + 1)\mathds{1}_2$. As opposed to the coherent case discussed above, in this case the displacement contribution~\eqref{F_x_u0} vanishes, and the QFI is fully determined by the covariance matrix term~\eqref{F_v_u0}.
Since $V_0$ is proportional to the identity, the only nonzero coefficients in Eqs.~\eqref{ab} are $a_2^2 = 8 (\partial_d N_0)^2$ and $b_2^2 = 16N_0^2\eta^2$, resulting in
\begin{equation}
     F_{d, {\rm th}} = \frac{(\partial_d N_0)^2}{N_0(N_0 +1)} + 4\eta^2 N_0.
    \label{F_v_uo_thermal}
\end{equation}
The $4\eta^2N_0$ term is identical to the one in Eq.~\eqref{F_x_uo_coherent}. Accordingly, when the mean photon number $N_0$ does not depend explicitly on the transverse displacement, we have the same QFI for thermal and coherent beams.
On the other hand, the explicit dependence of the mean photon number $N_0$ on the parameter induces a quite different dependence. 
To make this difference more explicit, we use $N_0 = |\alpha|^2$ to rewrite this term in function of the mean photon number $N_0$ also in the coherent case. Accordingly, we get $|\partial_d \alpha|^2 = (\partial_d N_0)^2/N_0$, which is a factor $N_0 + 1$ larger than the corresponding term in the thermal case.
As a consequence, if the $d-$dependence of mean photon number dominates the QFI, such as in the case of strong displacement-dependent losses, coherent states provide a significant advantage over thermal states. 
This is due to the fact that for coherent states, a variation of the mean photon number consists in a change of mean field, while for thermal states it is a change of the covariance matrix, and the former is more efficient than the latter in making two Gaussian distributions distinguishable.

\subsubsection{Populating the derivative mode}
It was demonstrated by~\cite{PinelPRA2012, gessner2022quantum}, that the QFI~\eqref{F_x_uo_coherent} can be enhanced by adding squeezing to the derivative mode $\partial_d u_0[d]({\bf r})$. In the following, we will see how our formalism recovers this result, to extend it to different states of the mode $u_0[d]({\bf r})$ and to take into account losses in the squeezed derivative mode.

When populating the derivative mode, the mode Hilbert space $\mathcal{H}_n$, as introduced in Sec.~\ref{Sec:mode_enc_param_est}, is spanned by $u_0[d]({\bf r}) = u({\bf r}- {\bf r}_0)$ and its normalized derivative $u_1[d]({\bf r}) = \partial_d u_0({\bf r}- {\bf r}_0)/\eta$. On the other hand, the mode Hilbert space $\mathcal{H}_\partial$ (see Sec.~\ref{Sec:mode_enc_param_est}) only contains the second derivative mode $u_2[d]({\bf r}) = \left(\partial_d u_1[d]({\bf r}) -\xi u_0[d]({\bf r})\right)/\zeta$, with $\xi = (\partial_d u_1| u_0)$ and $\zeta = ||\partial_d u_1[d]({\bf r}) -\xi u_0[d]({\bf r})|| $. Accordingly, we have 
\begin{align}
D_n &= \begin{pmatrix}
0 & \eta \mathds{1}_2 \\
\xi \mathds{1}_2 & 0
\end{pmatrix}, \quad
D_\partial = \begin{pmatrix}
0 \\
\zeta \mathds{1}_2 
\end{pmatrix}.
\end{align}
Furthermore, we assume that the derivative mode $u_1[d]({\bf r})$ has no mean field so that the mean field vector can be written as $\bar{\bf x}^\top= (q_0, p_0, 0, 0)$. Therefore, the mean field term of the QFI~\eqref{Fx} results in
\begin{equation}
F_{\bar{\bf x}} = {\bf y}^T V^{-1} {\bf y} = 
(\partial_d \bar{\bf x}_0^\top, \eta \bar{\bf x}_0^\top) V^{-1} \begin{pmatrix}
\partial_d \bar{\bf x}_0 \\\eta \bar{\bf x}_0 \\ 
\end{pmatrix}.
\label{Fx_u}
\end{equation}
As noted by~\cite{PinelPRA2012}, the QFI~\eqref{Fx_u} can be rewritten as a function of a unique element of the inverse covariance matrix $V_v^{-1}$ 
\begin{equation}
F_{\bar{\bf x}} = ||{\bf y}||^2(V^{-1}_v)_{0,0},
\label{F_x_squeezing}
\end{equation}
where $(V_v)_{0,0}$ is the variance of the $q-$quadrature of mode 
%\begin{subequations}
\begin{align}
v_0 [d] ({\bf r}, t) &= \frac{(\partial_d q_0 + i \partial_d p_0)}{||{\bf y}||}u_0[d] ({\bf r}, t) \label{v0}\\ 
&\quad +\frac{\eta (q_0+ip_0)}{||{\bf y}||}u_1[d] ({\bf r}, t) \nonumber
%\frac{(\partial_d q_0 + i \partial_d p_0)}{\sqrt{||\partial_d \bar{\bf x}_0 ||^2+\eta^2||\bar{\bf x}_0 ||^2}}u_0[d] ({\bf r}, t) \nonumber \\ &\;+ \frac{\eta (q_0+ip_0) }{\sqrt{||\partial_d \bar{\bf x}_0 ||^2+\eta^2||\bar{\bf x}_0 ||^2}}u_1[d] ({\bf r}, t) \label{v0}\\
%v_1 [d] ({\bf r}, t) &= -\frac{\eta (q_0 + i p_0)(\partial_d q_0 - i \partial_d p_0 )}{||\partial_d \bar{\bf x}_0||\sqrt{||\partial_d \bar{\bf x}_0 ||^2+\eta^2||\bar{\bf x}_0 ||^2}}u_0[d] ({\bf r}, t) \nonumber \\ &\;+ \frac{||\partial_d \bar{\bf x}_0||}{\sqrt{||\partial_d \bar{\bf x}_0 ||^2+\eta^2||\bar{\bf x}_0 ||^2}}u_1[d] ({\bf r}, t),
\end{align}
%\end{subequations}
%the QFI~\eqref{Fx_u} can be rewritten as a function of a unique element of the inverse covariance matrix $V_v^{-1}$ in this basis:
%\begin{equation}
%F_{\bar{\bf x}} = (V^{-1}_v)_{0,0}(||\partial_d \bar{\bf x}_0 ||^2+\eta^2||\bar{\bf x}_0 ||^2).
%\label{F_x_squeezing}
%\end{equation}
Accordingly, for states with a nonzero mean field and a parameter-independent covariance matrix (e.g. coherent states), it is necessary and sufficient to squeeze the $q$ quadrature of mode $v_0[d]({\bf r})$ to quantum enhance our beam positioning capability. 
It is interesting to observe that, if the mean field vector does not depend explicitly on the beam displacement $d$, i.e. $\partial_d q_0 = \partial_d p_0 = 0$, the mode $v_0[d]({\bf r})$ equals (up to a global phase) the derivative mode $u_1[d]({\bf r})$, which is orthogonal to the mode $u_0[d]({\bf r})$ that defines the beam shape. 
In this case, this effect has been exploited experimentally to enhance position estimation with a so called {\it quantum laser pointer} \citep{TrepsScience2003}.
In a more general scenario, e.g. in presence of position-dependent losses, Eq.~\eqref{v0} prescribes to squeeze a mode $v_0[d]({\bf r})$ which is partially overlapping with $u_0[d]({\bf r})$.

Let us now discuss how the covariance matrix term of the QFI is modified by population in the derivative mode $u_1[d]({\bf r})$. 
For simplicity, we will focus on the case where the population of mode $u_1[d]({\bf r})$ is fully uncorrelated with that of mode $u_0[d]({\bf r})$, therefore, the covariance matrix takes the block diagonal form 
\begin{equation}
    V = 
    \begin{pmatrix}
    V_0 & 0 \\
    0 & V_1
    \end{pmatrix}.
\end{equation}
Under these assumptions, the covariance matrix contribution to the QFI~\eqref{QFI_sigma} takes the form
\begin{align}
F_v &= \frac{1}{2}\sum_{l=0}^3  \left( \frac{a_l^2}{\nu_0^2 -(-1)^l} \right.\label{F_v_u1}\\ 
& \quad+  \left.\frac{\tilde{b}_l^2}{\nu_0 \nu_1 -(-1)^l} +  \frac{d_l^2}{\nu_1 -(-1)^l} \right),
\nonumber
\end{align}
where we have defined the coefficients
\begin{subequations}
\begin{align}
a_l^2 &= \tr \left[A_l S_0^{-1} (\partial_\vartheta V_0)  (S_0^\top)^{-1} \right]^2,  \\
\tilde{b}_l^2 &= \left(\eta \tr \left[A_l  S_1^{-1} \left( V_0 - \mathds{1}_{2}\right) \left(S_0^\top\right)^{-1} \right] \right. \\ &\quad + \left. \xi \tr \left[A_l  S_1^{-1} \left( V_1 - \mathds{1}_{2}\right) \left(S_0^\top\right)^{-1} \right]\right)^2 \nonumber  \\
&\quad+ \left(\eta \tr \left[A_l S_0^{-1} \left( V_0 - \mathds{1}_{2}\right) (S_1^{-1})^\top  \right]^2 \right. \nonumber \\  &\quad + \left. \xi \tr \left[A_l S_0^{-1} \left( V_1 - \mathds{1}_{2}\right) (S_1^{-1})^\top  \right]\right)^2, \nonumber\\ 
d_l^2 &= \zeta^2 \tr \left[A_l  S_1^{-1} \left( V_1 - \mathds{1}_{2}\right) \right]^2 \\&\quad+ \zeta^2 \tr \left[A_l \left( V_1 - \mathds{1}_{2}\right) (S_1^{-1})^\top  \right]^2.\nonumber 
\end{align}
\end{subequations}
%First, let us notice that if we assume mode $u_1[d]({\bf r})$ to be in vacuum, we have $V_1 = \mathds{1}_2$, implying $S_1 =  \mathds{1}_2$ and $\nu_1 = 1$. 
%Accordingly, we have that $\tilde{b}_l = b_l$ and $d_l =0$, so that Eq.~\eqref{F_v_u1} reduces to \eqref{F_v_u0} as it should be.

To further illustrate how to use Eq.~\eqref{F_v_u1} in practice, let us now consider the localisation of a thermal beam $u_0[d]({\bf r})$ aided by a squeezed vacuum state in the derivative mode $u_1[d]({\bf r})$. Accordingly, we have
\begin{align}
V_0 =  (2N_0 + 1)\mathds{1}_2, \quad
V_1 = \begin{pmatrix}
e^{-2r} & 0\\
0 & e^{2r}
\end{pmatrix},
\end{align}
which corresponds to
\begin{align}
S_0 = \mathds{1}_2, \quad {\rm with}  \quad
S_1 = \begin{pmatrix}
e^{-r} & 0\\
0 & e^{r}
\end{pmatrix},
\label{single_mode_S}
\end{align}
with $\nu_0 = 2N_0 + 1$ and $\nu_1 = 1$.
Under these assumptions, the only nonzero coefficients are $ a^2_2 = 2(2N_0^\prime+1)^2$, $\tilde{b}^2_1= 16(N_0\eta - \xi)^2\sinh^2 r$, $\tilde{b}^2_2 = 16N_0^2\eta^2\cosh^2r$, and $d^2_1= 2\zeta^2\sinh^2r$.
    Substituting into Eq.~\eqref{F_v_u1}, we obtain the following expression the QFI (for a zero mean state, the contribution in Eq.~\eqref{F_x_squeezing} vanishes)
\begin{align}
F_{d,{\rm th - sq}} &=\frac{(\partial_d N_0)^2}{N_0(N_0+ 1)} + \frac{4(N_0\eta -\xi)^2N_1}{N_0 + 1} \label{Fv-single-mode_squeez}\\
&\quad + 4N_0\eta^2(N_1+1) +4\zeta^2N_1, \nonumber
\end{align}
where we have introduced the number of photons $N_1 = \sinh^2 r$ in the squeezed derivative mode. 
Given that a thermal state has no preferred direction in phase space, we find that the result in Eq.~\eqref{Fv-single-mode_squeez} remains valid if we modify the squeezing direction. 
%We can verify that this is actually the case by replacing the matrix $S_1$ in Eq.~\eqref{single_mode_S} with $R(\theta)S_1R(\theta)^\top$, where $R(\theta)$ is an arbitrary rotation matrix.
Furthermore, we can see that the QFI~\eqref{Fv-single-mode_squeez} is always larger than the one in Eq.~\eqref{F_v_uo_thermal} for a thermal state alone.
This becomes particularly evident if we assume that $N_0$ does not explicitly depend on the parameter, and we consider the $N_0 \gg N_1 \gg 1$ limit, where we have $F_{d,{\rm th - sq}} \sim 4N_0(2N_1 +1)\eta^2 \sim 2 N_0 \eta^2 e^{2r} \sim e^{2r} F_{d,{\rm th}}/2$.

It is interesting to compare this result, with the quantum enhancement achievable with a coherent state in mode $u_0[d]({\bf r})$. For simplicity, let us consider the case where the mean field $\bar{\bf x}_0$ does not depend explicitly on the transverse beam displacement $d$. In such a case, combining Eq.~\eqref{F_x_squeezing} with Eq.~\eqref{Fv-single-mode_squeez} (setting the number of thermal photons to zero), we obtain
\begin{equation}
    F_{d,{\rm coh - sq}} = 4N_0\eta^2e^{2r} + 4(\xi^2 +\zeta^2)N_1.
\end{equation}
The second term is negligible for $N_0 \gg N_1$, and we obtain $ F_{d,{\rm coh - sq}} \sim e^{2r}F_{d,{\rm coh}}$.
Accordingly, for the positioning a bright thermal beam aided with a squeezed state in the derivative mode, we have a quantum enhancement which is just a factor two smaller than that we obtain adding squeezing in the derivative mode $u_1[d]({\bf r})$ when the mode $u_0[d]({\bf r})$ is in a coherent state.
We can understand this result by considering a thermal state as an ensemble average over coherent states with Gaussian distributed amplitudes and uniformly distributed phases. 
Accordingly, when adding squeezing in the derivative mode, the relative orientation between the coherent states in the ensemble and the squeezing will result sometimes in an enhancement and sometimes in a reduction of the sensitivity (see Eq.~\eqref{F_x_squeezing}). 
To make this statement more quantitative, we compute from Eqs.~\eqref{F_x_squeezing} and \eqref{Fv-single-mode_squeez} the average QFI of a coherent state in mode $u_0[d]({\bf r})$ combined with a squeezed vacuum state in the normalized derivative mode $u_1[d]({\bf r})$ with random, uniformly distributed squeezing directions
\begin{align}
    F_{d, {\rm avg}} &= \frac{4N_0\eta^2}{2\pi}\int ( e^{2r} \cos^2 \phi +  e^{-2r} \sin^2 \phi)d \phi \nonumber\\ & \quad + 4(\xi^2 +\zeta^2)N_1  \\ &
    = 4N_0\eta^2\cosh 2r + 4(\xi^2 +\zeta^2)N_1. \nonumber
\end{align}
While in general, the convexity of the QFI ensures $F_{d, {\rm avg}} \geq F_{d, {\rm th-sq}}$, for $N_0 \gg N_1 \gg 1$ we have $F_{d, {\rm avg}} \sim 2 N_0 \eta^2 e^{2r} \sim F_{d, {\rm th-sq}}$, which supports our interpretation that the the quantum advantage enabled by squeezing for thermal states can be seen as an average over the sensitivity enhancements/diminutions obtained for coherent states.

\begin{figure}
    \centering
    \includegraphics[width = \columnwidth]{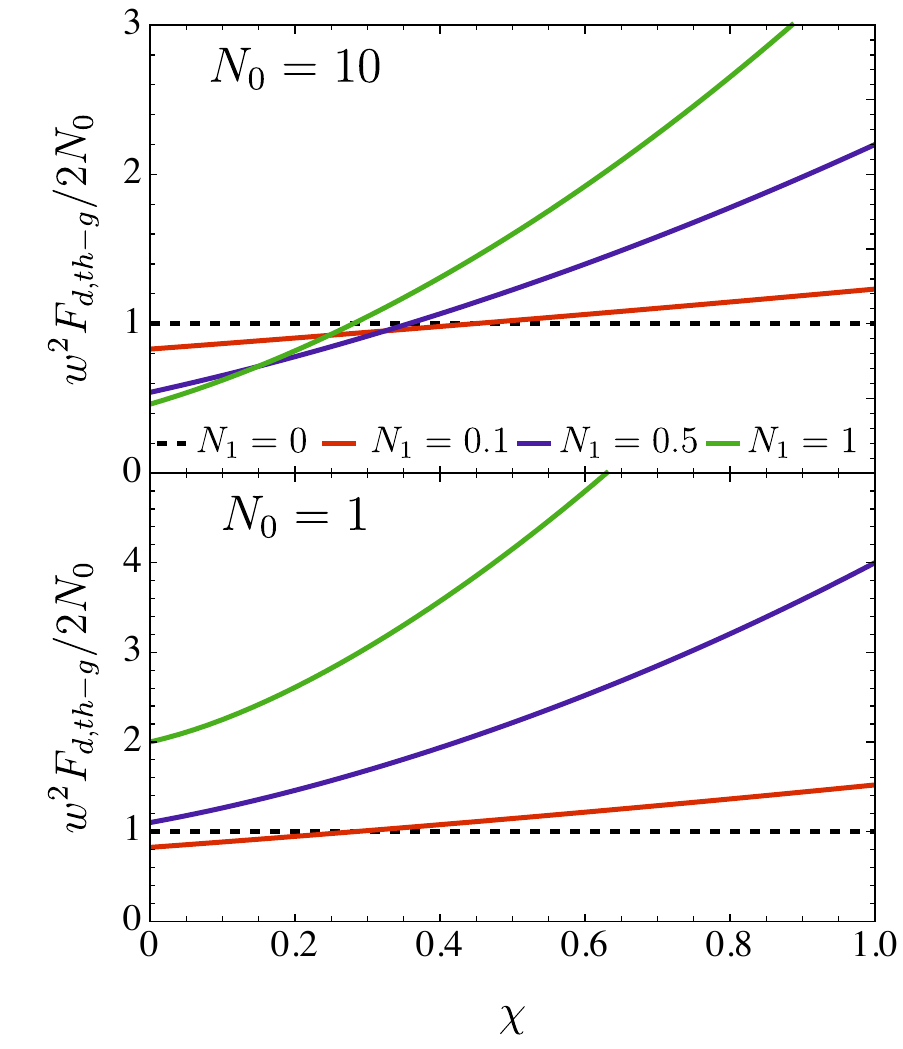}
    \caption{QFI (normalized by its maximum value when $u_1[d]({\bf r})$ is in vacuum) for the estimation of the transverse shift $d$ of a thermal Gaussian beam with mean photon number $N_0 = 10$ (top) and $N_0=1$ (bottom), assisted by a thermal squeezed state with mean photon number $N_1$, as function of the squeezing fraction $\chi$.}
    \label{fig:squeezing_fraction}
\end{figure}

%At this point, one might wonder if the increase in sensitivity observed above comes from the particular properties of the squeezed state we added to the normalized derivative mode $u_1[d]({\bf r})$, or if it is simply due to the fact that we are adding population to a mode that contains information about the parameter. 
We demonstrated above how squeezing in the normalized derivative mode $u_1[d]({\bf r})$ can lead to a sensitivity enhancement in the estimation of the displacement of a Gaussian beam. 
However, in practical situations it is hard to get a squeezed state which is not corrupted by noise. 
To illustrate what happens in these more practical scenarios, let us consider a thermal state in mode $u_0[d]({\bf r})$ and the derivative mode $u_1[d]({\bf r})$ populated with an arbitrary zero-mean Gaussian state, i.e. a squeezed thermal state.
Accordingly, we have
\begin{equation}
V_1 = (2N_T +1)\begin{pmatrix}
e^{-2r} & 0\\
0 & e^{2r}
\end{pmatrix},
\end{equation}
where $r$ quantifies the squeezing strength, while $N_T$ quantifies the amout of thermal noise.
Accordingly, the matrix $S_1$ in Eq.~\eqref{single_mode_S} remains the same, while the symplectic eigenvalue become $\nu_1 = 2N_T+1$.
The total photon number in the derivative mode for such a state is given by $N_1 = N_T + N_S +2N_TN_S$, with the squeezing contribution given by $N_S = \sinh^2 r$.
Following the same steps as above, we now obtain the following expression for the covariance matrix contribution to the QFI~\eqref{F_v_u1}
\begin{align}
&F_{d,{\rm th-g}}= \frac{(N_0^\prime)^2}{N_0(N_0+1)} + \frac{ 4 N_S \left(\xi (N_T+1) -\eta N_0\right)^2}{2 N_0 N_T+N_0+N_T+1} \nonumber \\
&+\frac{4 (N_S+1) \left(\eta ^2 N_0^2+N_0 N_T \left(2 \eta  \xi +\zeta ^2 (2 N_T+1)\right)\right)}{2 N_0 N_T+N_0+N_T}\nonumber \\
&+\frac{4 N_T^2 (N_S+1)\left(\zeta ^2+\xi ^2\right)}{2 N_0 N_T+N_0+N_T} + 4 N_S\zeta ^2 (N_T+1), \label{Fv-single-mode_general} 
\end{align}
which reduces to Eq.~\eqref{Fv-single-mode_squeez} when $N_T=0$.
On the other hand, when $N_S=0$ and the population of the derivative mode becomes purely thermal, we obtain
\begin{align}
F_{d,{\rm th-th}} &= \frac{(N_0^\prime)^2}{N_0(N_0+ 1)} + 4\frac{(\eta N_0+\xi N_1)^2}{2 N_0 N_1+N_0+N_1} \nonumber \\ 
&\quad +4\zeta ^2 N_1,
\label{Fv-single-mode_thermal}
\end{align}
and it is not hard to show that $F_{d,{\rm th-th}}$~\eqref{Fv-single-mode_thermal} is always smaller than the $F_{d,{\rm th-sq}}$~\eqref{Fv-single-mode_squeez}:
unsurprisingly, populating the derivative mode with squeezing is always better than populating that with thermal noise.
In fact, for small values of $N_1$, the QFI $F_{d,{\rm th-th}}$~\eqref{Fv-single-mode_thermal} is even smaller than that for an unpopulated derivative mode $F_{d,{\rm th}}$~\eqref{F_v_uo_thermal}.
To better illustrate this interplay between squeezing and thermal noise, we introduce the following parametrisation
\begin{align}
N_S = \chi N_1 \quad {\rm and} \quad N_T = \frac{(1-\chi)N_1 }{1+2\chi N_1},
\end{align}
which allows to vary the amount of squeezing and thermal noise while keeping constant the total number of photons $N_1$ in the derivative mode. 
In particular, for $\chi = 1$ the derivative mode is purely squeezed, while for $\chi = 0$ it is purely thermal, so that we can refer to $\chi$ as the {\it squeezing fraction}.
If we further assume that the beam we are trying to localize is Gaussian, i.e. $u({\bf r}) = \exp(-|{\bf r}|^2/2w^2)/\sqrt{\pi w^2}$, we show that for $\chi \geq 1/2$ and $N_1>0$, the QFI~\eqref{Fv-single-mode_general} is always larger than that for unpopulated derivative mode, i.e. for $N_1 =0$. 
On the other hand, as presented in Fig.~\ref{fig:squeezing_fraction}, for $\chi<1/2$ and small $N_1$ we obtain a worse sensitivity compared to that when the derivative mode is in vacuum.   

Optical metrology protocols are generally very sensitive to photon losses, it is therefore important to illustrate how such losses can be taken account.
Accordingly, it is useful to note that for a thermal state of mode $u_0[d]({\bf r})$, and an arbitrary zero-mean Gaussian state of the derivative mode $u_1[d]({\bf r})$, the QFI maintains the form~\eqref{Fv-single-mode_general} even after losses. 
In fact, it is sufficient to perform the following substitutions
\begin{align}
N_0 &= N_0^{\rm in}\kappa_0 \\
N_T &= \frac{1}{2}\left(\left[\left(2\kappa_1(2N_S^{\rm in}N_T^{\rm in}+N_T^{\rm in}+N_S^{\rm in})+1\right)^2 \right. \right. \\&\quad \left. \left. -4N_S^{\rm in}(2N_T^{\rm in}+1)^2(N_S^{\rm in}+1)\right]^{1/2} -1 \right) \nonumber \\ 
\sinh (2 r) & = \kappa_1 \frac{2N_T^{\rm in}+1}{2N_T+1} \sinh (2 r^{\rm in}),
\end{align}
where $\kappa_0$ and $\kappa_1$ are the attenuation coefficients of the two modes $u_0[d]({\bf r})$, and $u_1[d]({\bf r})$, respectively; while $N_0^{\rm in}$, $N_T^{\rm in}$ and $N_S^{\rm in} = \sinh^2 r^{\rm in}$ are the populations of the mode $u_0[d]({\bf r})$, and the thermal and squeezing components of the population of the mode $u_1[d]({\bf r})$, respectively.
Finally, in some applications, the attenuation coefficients $\kappa_0$ and $\kappa_1$ can be parameter dependent. 
In those cases, not only $N_0$ depends on the transverse displacement $d$ (as taken into account by the first term in Eq.~\eqref{Fv-single-mode_general}), but also $N_T$ and $r$. 
This leads to an additional term in the QFI which takes the form
\begin{align}
&\qquad\frac{1}{2}\sum_{l=0}^3\frac{\tr \left[A_l S_1^{-1} \partial_d V_1 \left(S_1^{-1}\right)^\top \right]^2 }{\nu_1^2 - (-1)^l} \\
&\quad =\frac{\left((2 N_T+1) (\partial_d r) \cosh r-2 (\partial_d N_T)\sinh r \right)^2}{4 N_T (N_T+1)+2} \nonumber\\ & \qquad +\frac{\left((2 N_T+1) (\partial_d r) \sinh r-2 (\partial_d N_T) \cosh r \right)^2}{4 N_T(N_T+1)}. \nonumber
\end{align}

\subsection{Temporal separation between pulses}
\begin{figure}[htb!]
 \begin{tikzpicture}
    \node at (0.89,5.5) {\includegraphics[width=0.5\columnwidth]{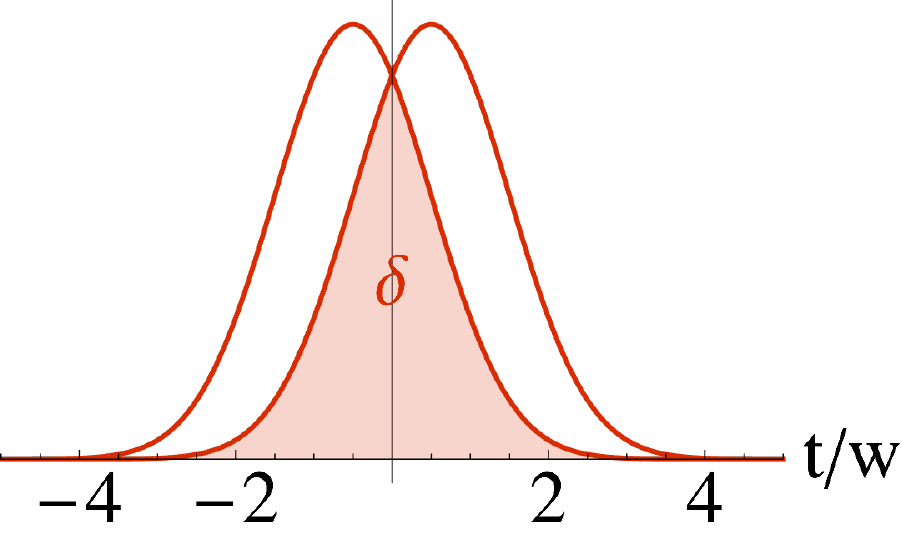}};
    \node at (0,0) {\includegraphics[width=\columnwidth]{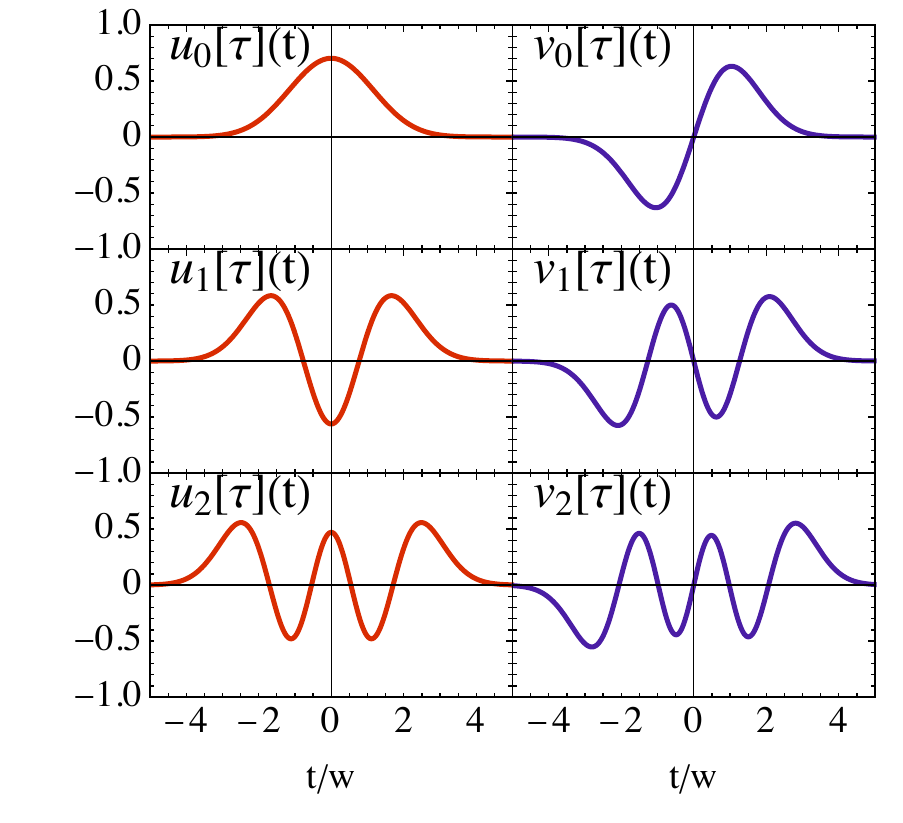}};
    \end{tikzpicture}
    \caption{(top) Two Gaussian pulses temporally separated by $\tau = w$, with their overlap $\delta$ represented as a shaded area. (bottom) Set of orthonormal modes constructed from the two Gaussian pulses and their first and second derivatives with respect to $\tau$. All modes have definite parity, in particular the modes $u_i[\tau](t)$ (left - red) are even functions, while the modes $v_i[\tau](t)$ (right - blue) are odd functions.}
    \label{fig:modes}
\end{figure}

As a second example, we consider the estimation of the time delay $\tau$ between two light pulses with the same temporal profile defined by the mode $u(t)$,
which for simplicity, we will assume to be real and even, i.e. $u(t) = u(-t)$.
From a parameter estimation point of view, this problem is most interesting when the separation $\tau$ between the pulses is smaller than (or comparable to) the pulse width. 
In this context, there is a finite overlap between the modes $u(t-\tau/2)$ and $u(t+\tau/2)$ (see Fig.~\ref{fig:modes})
\begin{equation}
    \delta = \int u(t-\tau/2) u(t+\tau/2) d t.
    \label{delta}
\end{equation}
Accordingly, as discussed by \cite{LupoPirandola,Sorelli_2021_letter,Sorelli_2021_long} for the spatial domain, it is convenient to describe the problem in terms of the two orthonormal modes 
\begin{subequations}
\begin{align}
u_0[\tau] (t) &= \frac{u(t-\tau/2) + u(t+\tau/2) }{\sqrt{2(1+\delta)}}, \\
v_0[\tau] (t) &= \frac{u(t-\tau/2) - u(t+\tau/2) }{\sqrt{2(1-\delta)}}. 
\end{align}
\label{u0v0}
\end{subequations}
We are interested in computing the QFI for the estimation of $\tau$, when the two modes~\eqref{u0v0}, and eventually their derivatives, are populated. 
Accordingly, we complement the modes~\eqref{u0v0} with their orthonormalized first and second derivatives (see App.~\ref{App:upmd} for detailed calculations):
\begin{subequations}
\begin{align}
    u_1[\tau](t) &= \partial_\tau u_0[\tau](t)/\eta_u \\
    v_1[\tau](t) &= \partial_\tau v_0[\tau](t)/\eta_v \\
    u_2[\tau](t) &= (\partial_\tau u_1[\tau](t) - \xi_u u_0[\tau](t))/\zeta_u \\ 
    v_2[\tau](t) &= (\partial_\tau v_1[\tau](t) - \xi_v v_0[\tau](t))/\zeta_v,
\end{align}
\label{u1v1u2v2}
\end{subequations}
where $\eta_u = ||\partial_\tau u_0[\tau](t)||$, $\eta_v = ||\partial_\tau v_0[\tau](t)||$, $\xi_u = (\partial_\tau u_1| u_0)$, $\xi_v = (\partial_\tau v_1| v_0)$, $\zeta_u = ||\partial_\tau u_1[\tau](t) - \xi_u u_0[\tau](t)||$ and $\zeta_u = ||\partial_\tau v_1[\tau](t) - \xi_v v_0[\tau](t)||$.
The shapes of the modes $u_i[\tau](t)$ and $v_i[\tau](t)$ for the specific case of Gaussian pulses $u(t) = e^{-t^2/2 w^2}/(\pi w^2)^{1/4}$ are presented in Fig.~\ref{fig:modes}.
Using the modes~\eqref{u0v0} and \eqref{u1v1u2v2}, we can express the matrices $D_n$ and $D_\partial$ (see Sec.~\ref{sec:covariance_mat_contr}) as 

\begin{align}
D_n  &= 
\begin{pmatrix} 
0 & D_\eta\\
D_\xi & 0 
\end{pmatrix}, \quad D_\partial = \left(0, 0,D_\zeta\right)^\top, \quad {\rm with}  \nonumber\\ 
D_\eta &= \begin{pmatrix}
\eta_u\mathds{1}_2 & 0\\
0 & \eta_v\mathds{1}_2 
\end{pmatrix}, \; 
D_\xi = \begin{pmatrix}
\xi_u \mathds{1}_2 & 0\\
0 & \xi_v \mathds{1}_2 
\end{pmatrix}, \nonumber \\
D_\zeta &= \begin{pmatrix}
\zeta_u \mathds{1}_2 & \zeta_v \mathds{1}_2 
\end{pmatrix}. 
\end{align}

We have now specified all the mode-related quantities needed to compute the QFI for the estimation of the temporal separation $\tau$ between pulses. 
To proceed, we now make some further assumptions on the quantum state of the pulses. 
In particular, we consider the modes $u_0[\tau](t)$ and $v_0[\tau](t)$ to be in a general Gaussian state, and we allow for auxiliary population of the orthogonalized first derivative modes $u_1[\tau](t)$ and $v_1[\tau](t)$ with no mean field\footnote{Note that we allow for correlations (classical or quantum) between the modes $u_0[\tau](t)$ and $v_0[\tau](t)$. To focus on the role of squeezing in the derivative modes, in the examples contained this paper, we do not consider such correlations. However, we studied their role in diffraction-limited imaging in \cite{Sorelli_PRRes_2022}.}. 
Accordingly, we can write the mean field vector as $\bar{\bf x}^\top = (\bar{\bf x}^\top_0, 0)$, with $\bar{\bf x}_0 = (q_{u_0}, p_{u_0},q_{v_0},p_{v_0})$.
The mean field term of the QFI~\eqref{Fx} can then be expressed as 
\begin{align}
F_{\bar{\bf x}} = (\partial_\tau \bar{\bf x}_0, D_n \bar{\bf x}_0)^\top V^{-1} (\partial_\tau \bar{\bf x}_0, D_n \bar{\bf x}_0)
\label{Fx_tau}
\end{align}
Similarly to what we discussed in Sec.~\ref{sec:spatial_positioning}, for every state with a nonzero mean field, i.e. with $||\bar{\bf x}_0|| \neq 0$, there always exists a mode basis where $(\partial_\tau \bar{\bf x}_0, D_n \bar{\bf x}_0)$ has only one nonzero component.
In other words, there always exists an orthogonal transformation $O$ s.t. $O(\partial_\tau \bar{\bf x}_0, D_n \bar{\bf x}_0) = (\sqrt{||\partial_\tau \bar{\bf x}_0||^2+ ||D_\eta \bar{\bf x}_0||^2}, 0, 0, 0)$. Accordingly the QFI only depends on the inverse covariance matrix element $\left(O^T V^{-1} O\right)_{0,0}$.
Therefore, the use of quantum resources, such as squeezing, to increase such matrix element can lead to an enhanced sensitivity \citep{PinelPRA2012}.
%It is interesting to notice that if we assume that our pulses have a mean field that does not depend explicitly on their temporal separation $\partial_\tau \bar{\bf x}_0 = 0$, we have $F_{\bar{\bf x}} = \bar{\bf x}_0^\top D_\eta V_1^{-1} D_\eta  \bar{\bf x}_0$, with $V_1$ the block of the covariance matrix $V$ associated with the normalized derivative modes $u_1[\tau](t)$ and  $v_1[\tau](t)$.
%Accordingly, by squeezing an appropriate linear combination of these derivative modes the mean field contribution to the QFI can be cast in the form $F_{\bar{\bf x}} =||D_\eta \bar{\bf x}_0||^2 e^{2r}$.

Let us now have a look at the covariance matrix contribution to the QFI~\eqref{QFI_sigma}. 
To this goal, we will make the simplifying assumption that the population of the derivative modes $u_1[\tau](t)$ and  $v_1[\tau](t)$ is uncorrelated with that of the symmetric and antisymmetric superpositions $u_0[\tau](t)$ and  $v_0[\tau](t)$ of the pulses we want to separate, so that we can write the covariance matrix in block diagonal form
\begin{equation}
    V = \begin{pmatrix}
    V_0 & 0 \\
    0 & V_1
    \end{pmatrix}.
    \label{V0V1}
\end{equation}
Under these assumptions, the covariance matrix contribution to the QFI takes the form 
\begin{align}
F_V = &\frac{1}{2} \sum_{l=0}^3 \sum_{jk=0,1}\left( \frac{\left(a_l^{jk} \right)^2}{\nu_0^j \nu_0^k - (-1)^l} + \frac{\left(b_l^{jk} \right)^2}{\nu_0^j \nu_1^k - (-1)^l} \right. \label{F_v_uv}\\
& \left. \quad+  \frac{\left(c_l^{jk} \right)^2}{\nu_1^j \nu_1^k - (-1)^l} + \frac{\left(d_l^{jk} \right)^2}{ \nu_1^j - (-1)^l}\right), \nonumber
\end{align}
where we introduced the coefficients
\begin{subequations}
\begin{align}
\left(a_l^{jk} \right)^2 &= \left(\tr \left[ A^{(jk)}_l \partial_\tau V_0\right] \right)^2, \\
\left(b_l^{jk} \right)^2 &=  \left( \tr \left\{  A_l^{(jk)} S_0^{-1}\left[ D_\xi^T (V_1 - \mathds{1}_4) \right. \right. \right. \\ &\quad+\left. \left. \left. (V_0 - \mathds{1}_4)D_\eta \right] \left(S_1^{-1}\right)^\top\right\} \right)^2 \nonumber\\
& \quad + \left( \tr \left\{  A_l^{(jk)} S_1^{-1}\left[ D_\xi^T (V_1 - \mathds{1}_4)  \right. \right. \right. \nonumber\\ &\quad +\left. \left. \left. (V_0 - \mathds{1}_4)D_\eta \right] \left(S_0^{-1}\right)^\top\right\} \right)^2, \nonumber\\
\left(c_l^{jk} \right)^2 &= \left(\tr \left[ A^{(jk)}_l \partial_\tau V_1\right] \right)^2, \\
\left(d_l^{jk} \right)^2 &=  \left( \tr \left\{A_l^{(jk)} D_\zeta (V_1 - \mathds{1}_4) \left(S_1^{-1}\right)^\top \right\} \right)^2,
\end{align}
\label{abcd}
\end{subequations}
with the matrices $A_l^{(jk)}$ as defined in Sec.~\ref{sec:Q_est_theory} and $\nu_0^j$ ($\nu_1^j$) the symplectic eigenvalues of the covariance matrix $V_0$ ($V_1$).
Accordingly, we have four groups of addends in the QFI~\eqref{F_v_uv}: The first one, depending on the coefficients $a_l^{jk}$, describes the contribution of the population of the symmetric $u_0[\tau](t)$ and antisymmetric $v_0[\tau](t)$ superpositions of the two pulses. These terms are nonzero if and only if the covariance matrix $V_0$ explicitly depends on the temporal separation, i.e. $\partial_\tau V_0 \neq 0$. Similarly, the third group of addends, depending on the coefficients $c_l^{jk}$, takes into account the population of the derivative modes $u_1[\tau](t)$ and $v_1[\tau](t)$, and is nonzero if and only if $\partial_\tau V_1 \neq 0$.
The second group of addends, containing the coefficients $b_l^{jk}$, takes into account how variations of the temporal separation $\tau$ leads to coupling between the modes $u_0[\tau](t)$, $v_0[\tau](t)$ and their derivative $u_1[\tau](t)$, $v_1[\tau](t)$. 
Finally, the addends containing the coefficients $d_l^{jk}$ take into account how due to variations of $\tau$ the derivative modes $u_1[\tau](t)$, $v_1[\tau](t)$ couple to the second derivative modes $u_2[\tau](t)$ and $v_2[\tau](t)$.

\begin{figure}[htb!]
    \centering
    \includegraphics[width = \columnwidth]{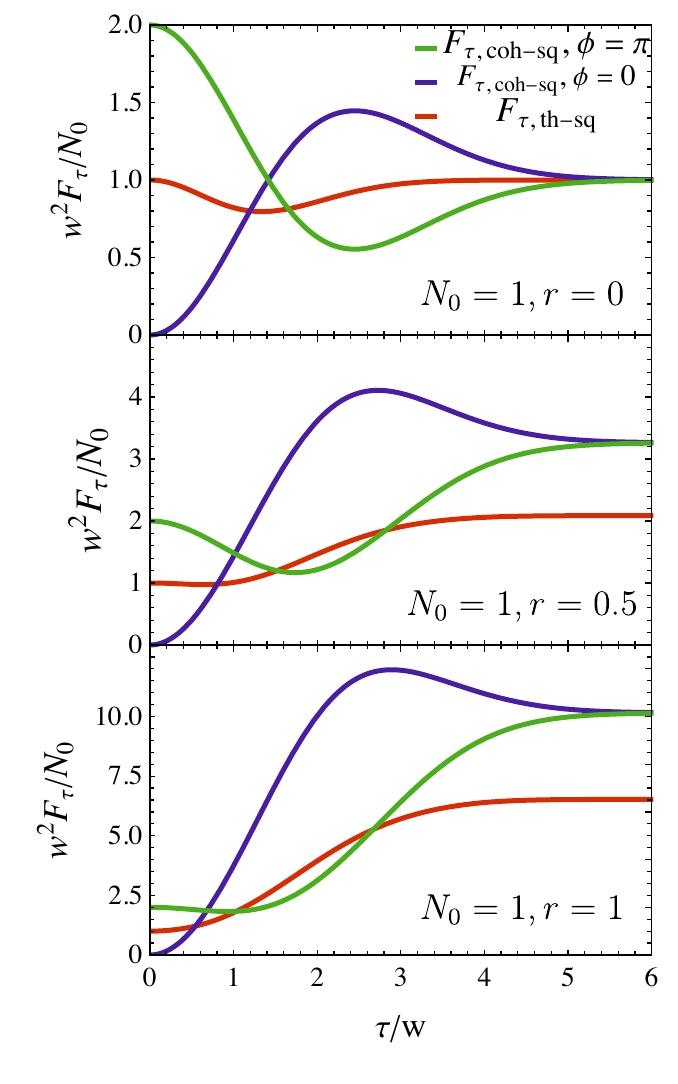}
    \caption{QFI (Normalized to its maximum value for thermal states in modes $u_0[\tau](t)$ and $v_0[\tau](t)$) for the estimation of the temporal separation $\tau$ between two thermal (red) or coherent pulses, either in phase (blue) or out of phase (green), as a function of the temporal separation $\tau$ in units of the pulse width $w$. 
    For each panel, we considered a mean photon number of $N_0 = 1$ per pulse, and different levels of squeezing in the derivative modes, as quantified by the parameter $r = 0$ (top), $r = 0.5$ (middle), and $r = 1$ (bottom). 
    The pulse shape is assumed Gaussian $u(t) = e^{-t^2/2w^2}/(\pi w^2)^{1/4}$ for all panels.}
    \label{fig:QFI_tau}
\end{figure}

Let us now evaluate the QFI~\eqref{F_v_uv} for a specific quantum state of the two pulses. 
In particular, we are interested in two equally-bright fully-incoherent pulses whose intensity distribution is given by 
\begin{align}
    I(t) &= \langle \hat{E}^\dagger(t) \hat{E}(t) \rangle \nonumber \\
    &= N_0 (|u(t-\tau/2)|^2 + |u(t+\tau/2)|^2),
    \label{intesity}
\end{align}
where we introduced the mean number of photons per pulse $N_0$, and the electric field operator $\hat{E}(t) = \sum_j \left(\hat{a}_j u_j[\tau](t) + \hat{b}_j v_j[\tau](t) \right)$, with $\hat{a}_j$ and $\hat{b}_j$ the annihilation operators associated with the even and odd modes $u_j[\tau](t)$ and $v_j[\tau](t)$, respectively (see Fig.~\ref{fig:modes}).
It is not hard to see that the intensity distribution $I(t)$~\eqref{intesity} is achieved by a thermal state of the modes $u_0[\tau](t)$ and $v_0[\tau](t)$, with mean photon numbers $N_u = N_0(1+\delta)$ and $N_v = N_0(1-\delta)$, respectively.
Such a state has no mean field $\bar{\bf x}_0 = 0$, so that its QFI is fully determined by Eq.~\eqref{F_v_uv}, and has a covariance matrix
\begin{equation}
    V_0 = \begin{pmatrix}
    \left(2N_0(1+\delta) + 1\right)\mathds{1}_2 & 0 \\
    0 &  \left(2N_0(1-\delta) + 1\right)\mathds{1}_2
    \end{pmatrix}.
    \label{V_0_tau}
\end{equation}
In Sec.~\ref{sec:spatial_positioning}, we have seen that adding squeezing to the derivative mode improve the sensitivity, even for the spatial localization of an incoherent thermal beam. 
To verify, whether this is the case also for the temporal separation between two thermal pulses, we assume the derivative modes $u_1[\tau](t)$ and $v_1[\tau](t)$ to be populated by two independent, equally-squeezed vacuum states, described by the covariance matrix
\begin{equation}
    V_1 = \begin{pmatrix}
    e^{-2r} & 0 & 0 & 0 \\
    0 & e^{2r}& 0 & 0 \\
    0 & 0 & e^{-2r} & 0\\
    0 & 0 & 0 &  e^{2r} \\
    \end{pmatrix}.
    \label{V_1_tau}
\end{equation}
For such a quantum state, the QFI~\eqref{F_v_uv} takes the form (see App.~\ref{App:QFI_tau_th_squ} for the explicit calculation of the coefficients~\eqref{abcd}) 
\begin{align}
F_{\tau,{\rm th-sq}}  &= \frac{2 N_0 \left[1+ N_0 (1+\delta^2) \right] \left(\partial_\tau \delta \right)^2}{(1-\delta^2)\left[(1+ N_0)^2 - (N_0 \delta)^2 \right]} \label{QFI_tau_therm_squeez} \\ 
&\quad +2(\zeta_u^2 + \zeta_v^2)\sinh^2 r \nonumber\\
&\quad + 4 N_0 \left[\eta^2_u (1+\delta) +\eta_v^2(1-\delta)\right]\cosh^2 r \nonumber \\
&\quad + 4\frac{\left( N_0(1+\delta)\eta_u - \xi_u \right)^2}{1+ N_0(1+\delta)} \sinh^2 r \nonumber \\ 
&\quad + 4\frac{\left( N_0(1-\delta)\eta_v - \xi_v \right)^2}{1+ N_0(1 - \delta)}\sinh^2 r. \nonumber
\end{align}
The behavior of the QFI~\eqref{QFI_tau_therm_squeez}, for Gaussian pulses, is plotted as red lines in Fig.~\ref{fig:QFI_tau}.

For comparison, we will now also evaluate the QFI for the temporal separation of two equally bright fully coherent pulses. 
As opposed to Eq.~\eqref{intesity}, in this case the intensity distribution also contains an interference term depending on the relative phase $\phi$ between the coherent pulses
\begin{align}
    I(t) &= \langle \hat{E}^\dagger(t) \hat{E}(t) \rangle \nonumber \\
    &= N_0 (|u(t-\tau/2)|^2 + |u(t+\tau/2)|^2)  \label{intesity_coherent}\\
    &\quad + 2 N_0 u(t-\tau/2)u(t+\tau/2) \cos \phi. \nonumber
\end{align}
Such an intensity distribution can be obtained by populating the modes $u_0[\tau](t)$ and $v_0[\tau](t)$ with coherent states, whose covariance matrix is the identity $V_0 = \mathds{1}_4$, and whose mean field is given by $\bar{\bf x}_0 = (\bar{\bf x}_u, \bar{\bf x}_v)^\top$, with
\begin{subequations}
\begin{align}
    \bar{\bf x}_u & = \sqrt{2N_0(1+\delta)}(1+\cos \phi, \sin \phi) \\
    \bar{\bf x}_v & = \sqrt{2N_0(1-\delta)}(1-\cos \phi, -\sin \phi).
\end{align}
\label{xuxv}
\end{subequations}
In particular, from Eqs.~\eqref{xuxv}, we can see that for in-phase ($\phi = 0$) coherent pulses, the mean field is fully determined by the $q$ quadrature of mode $u_0[\tau](t)$.
Similarly, when the two coherent pulses are out of phase ($\phi = \pi$) the mean field is fully determined by the $q$ quadrature of mode $v_0[\tau](t)$.
As we did for thermal sources, we are going to consider the derivative modes $u_1[\tau](t)$ and $v_1[\tau](t)$ by two independent squeezed vacuum states (see Eqs.~\eqref{V0V1} and \eqref{V_1_tau}).
Under these assumptions, the covariance matrix contribution to the QFI can be obtained simply by setting $N_0 = 0$ into Eq.~\eqref{QFI_tau_therm_squeez}.
Accordingly, we have
\begin{equation}
    F_{\tau, {\rm coh-sq}} =  F_{\bar{\bf x}, {\rm coh-sq}} + \left. F_{\tau,{\rm th-sq}}\right\vert_{N_0 =0},
    \label{QFI_tau_coh_sq}
\end{equation}
where the displacement term $F_{\bar{\bf x}, {\rm coh-sq}}$ can be computed from Eq.~\eqref{Fx_tau}, and reads
\begin{align}
    F_{\bar{\bf x}, {\rm coh-sq}} & = (\partial_\tau \bar{\bf x}_0)^\top V_0^{-1} (\partial_\tau \bar{\bf x}_0) + \bar{\bf x}_0^\top D_\eta V_1^{-1} D_\eta \bar{\bf x}_0 \nonumber \\
    & = \frac{2N_0(1-\delta\cos\phi)(\partial_\tau \delta)^2}{1-\delta^2} \label{F_x_tau_coh}  \\
    &\quad+ 2 N_0 e^{2r}\eta_u^2(1+\delta)(1+\cos\phi)^2 \nonumber\\ 
    &\quad+  2 N_0 e^{2r}\eta_v^2(1-\delta)(1-\cos\phi)^2 \nonumber\\
    &\quad+ 2 N_0 e^{-2r}  \sin^2 \phi [\eta_u (1+\delta) +\eta_v (1-\delta)]. \nonumber
\end{align}
We can see that, when the two coherent pulses are either in phase ($\phi = 0$) or out of phase ($\phi = \pi$), the last line in Eq.~\eqref{F_x_tau_coh} vanishes and only the squeezing-enhanced term proportional to $e^{2r}$ survives. 
This is consistent with the fact that the covariance matrix $V_1$~\eqref{V_1_tau} presents squeezing along the $q$ quadrature of modes $u_1[\tau](t)$ and $v_1[\tau](t)$, and for $\phi = 0, \pi$ the mean field~\eqref{xuxv} has a vanishing $p$ quadrature.
The QFI for the separation $\tau$ between in phase and out of phase coherent pulses are presented as blue and green lines in Fig.~\ref{fig:QFI_tau}.

Let us now compare the expressions for the QFI for thermal and coherent pulses aided by squeezing in the derivative modes reported in Eqs.~\eqref{QFI_tau_therm_squeez} and \eqref{QFI_tau_coh_sq}, respectively.
We start by comparing the behaviours for vanishingly small separations $\tau \to 0$. 
In this regime (see App.~\ref{App:upmd}), we have $(\partial_\tau \delta)^2/(1-\delta^2) \sim  (\Delta k)^2 $, with 
\begin{equation}
    (\Delta k)^2 = \int [\partial_t u(t)]^2 dt,
\end{equation}
and $\eta_u \sim \eta_v \sim \xi_u \sim \xi_v \sim \zeta_u \sim \zeta_v \sim 0$, which implies
\begin{subequations}
\begin{align}
    F_{\tau, {\rm th-sq}}  &\xrightarrow{\tau \to 0} 2 N_0 (\Delta k)^2, \\
    F_{\tau, {\rm coh-sq}} &\xrightarrow{\tau \to 0} 2 N_0 (\Delta k)^2 (1 + \cos \phi).
\end{align}
\end{subequations}
Accordingly, independently of the squeezing value $r$, the QFI for in phase coherent pulses vanishes for $\tau \to 0$, while that for out of phase coherent pulses is twice the one for incoherent pulses (see Fig.~\ref{fig:QFI_tau} where for Gaussian pulses we have $(\Delta k)^2 = 1/2w^2$).

To better understand this behavior, let us recall that the quantum state of the finite overlap $\delta$ between the two pulses induces a $\tau-$dependent population of the symmetric and antisymmetric modes $u_0[\tau](t)$ and $v_0[\tau](t)$. 
It is this dependence on temporal separation, which enters the QFI through $\partial_\tau \bar{\bf x}_0$ (in the coherent case) and $\partial_\tau V_0$ (in the incoherent case), that dominates the QFI behavior for $\tau \to 0$.
This implies that the population of the derivative modes $u_1[\tau](t)$ and $v_1[\tau](t)$, and in the particular the squeezing thereof, has no impact on the $\tau \to 0$ behavior of the QFI.

On the contrary, for separations much larger than the pulses' width, i.e. $\tau \Delta k\gg 1$, the overlap $\delta$ tends to zero, and the populations of the modes $u_0[\tau](t)$ and $v_0[\tau](t)$ become parameter independent.
The QFI is then dominated by the noise in derivative modes $u_1[\tau](t)$ and $v_1[\tau](t)$. In particular, we have 
\begin{subequations}
\begin{align}
    &F_{\tau, {\rm th-sq}} \xrightarrow{\tau \to \infty} 2(\Delta k)^2 \left(2 \sinh^2 r + N_0 \cosh 2r \right), \\
    &F_{\tau, {\rm coh-sq}} \vert_{\phi =0, \pi} \xrightarrow{\tau \to \infty} 2(\Delta k)^2 \left(2 \sinh^2 r + N_0 e^{2r} \right).
\end{align}
\end{subequations}
Accordingly, for large temporal separations $\tau$ we have a squeezing enhancement. 
Such an enhancement is always larger for coherent pulses than for thermal pulses. 
However, similarly to what we observed for the spatial localization of a beam, the QFI enhancement for large $\tau$ in the coherent case is at most a factor two larger than that in the thermal one.

\section{Conclusion}
\label{Sec:conclusion}
In this paper, we determined the ultimate sensitivity limit for the estimation of a parameter encoded into the quantum state as well as the mode structure of a multimode Gaussian state of the electromagnetic field.
In particular, we presented an analytical expression for the QFI, bounding the estimation sensitivity through the Cram\'er-Rao lower bound, which can be calculated from the first two moments of the states and the dependence on the parameter of the mode functions.
Such an expression expands the field of use of Gaussian quantum metrology to the estimation of parameters encoded into the spatio-temporal distribution of an electromagnetic signal.
We illustrated how to apply our general formalism by studying two paradigmatic problems: the estimation of the transverse displacement of a beam, and of the temporal separation between two pulses.

In the study of the transverse displacement we showed that if the mean photon number of the beam is independent of its transverse position, the displacement of a coherent and thermal beam can be estimated with the same sensitivity.
On the other hand, if the mean number of photons $N_0$ in the beam depends on its transverse displacement, e.g. because of position dependent losses, this dependence adds an additional term to the QFI which is $\sim N_0$ times larger for coherent beams than for thermal ones.
Furthermore, we showed that the sensitivity in the estimation of a transverse displacement can be enhanced by adding squeezing to a mode shaped like the derivative of the beam. Such a squeezing-enabled quantum enhancement is at most a factor two larger for coherent beams than for thermal ones.

We then moved to the time domain and considered the estimation of the temporal separation between two coherent or thermal pulses.
Such pulses are described by two temporal modes (the symmetric and anti-symmetric superpositions of the pulses) whose shape and populations depend on the separation parameter $\tau$. 
We showed that the interplay between these two dependences plays a fundamental role in the choice of which modes one needs to squeeze to achieve a quantum enhancement.
For large temporal separations, when the pulses have a negligible overlap, they are are most sensitive to the changes in the mode shapes. 
Accordingly, in this regime a quantum enhancement is possible by adding squeezing to the derivatives of the symmetric and anti-symmetric superpositions of the pulses.
As for the transverse displacement estimation, the quantum enhancement achieved for coherent pulses is at most a factor two larger than the one obtained for thermal ones.
On the other hand, for small temporal separations, when the pulses have a significant overlap, the QFI is dominated by how photons redistributes among the symmetric and anti-symmetric superpositions of the pulses. 
As a consequence, populating the derivative modes has no effect on the sensitivity in this regime. 

Our approach could be readily applied to other mode-encoded parameter estimation scenarios in various field of science and technology ranging from astronomy to microscopy \citep{gessner2022quantum}.
Moreover, parameters encoded into time-frequency modes appears in the characterization of frequency combs \citep{CaiNPJquantum2021}, or in radars that estimate the distance of a reflecting target from the temporal profile of chirped pulses \citep{vanTrees2001_I,vanTrees2001_III} (recent studies have addressed this problem in the quantum regime \citep{ZhuangPRL2022,gessner2022quantum}). 
Finally, the applicability of our approach could be further broadened by considering the simultaneous estimation of multiple parameters \citep{NicholsPRA2018}. 

\acknowledgments
We are very grateful to Claude Fabre for our illuminating discussions on the role of modes in quantum optical metrology, which have been a main source of inspiration for this work.
We also thank Ilya Karuseichyk for useful discussion.
This work was partially funded by French ANR under COSMIC project (ANR-19-ASTR0020-01). This work received funding from the European Union’s Horizon 2020 research and innovation programme under Grant Agreement No. 899587. 
This work was carried out during the tenure of an ERCIM ‘Alain Bensoussan’ Fellowship Programme.
This work was funded by MCIN/AEI/10.13039/501100011033 and the European Union “NextGenerationEU” PRTR fund [RYC2021-031094-I]. This work has been founded by the Ministry of Economic Affairs and Digital Transformation of the Spanish Government through the QUANTUM ENIA project call - QUANTUM SPAIN project, by the European Union through the Recovery, Transformation and Resilience Plan - NextGenerationEU within the framework of the Digital Spain 2026 Agenda, and by the CSIC Interdisciplinary Thematic Platform (PTI+) on Quantum Technologies (PTI-QTEP+).

%This work received funding from Ministerio de Ciencia e Innovaci\'{o}n (MCIN) / Agencia Estatal de Investigaci\'on (AEI) for Project No. PID2020-115761RJ-I00 and support of a fellowship from ``la Caixa” Foundation (ID 100010434) and from the European Union’s Horizon 2020 research and innovation program under Marie Sk\l{}odowska-Curie Grant Agreement No. 847648, fellowship code LCF/BQ/PI21/11830025. This work was partially funded by CEX2019-000910-S [MCIN/AEI], Fundaci\'o Cellex, Fundaci\'o Mir-Puig, and Generalitat de Catalunya through CERCA.

\appendix
\section{Properties of the basis-change matrices}
\label{App:OnON}
Here we derive some useful properties of the matrix $O$, and its blocks $O_n$ and $O_N$ from the properties of the $n$ initially populated modes $\{u_k[\vartheta]({\bf r}, t)\}$ and their derivatives $\{\partial_\vartheta u_k[\vartheta]({\bf r}, t)\}$.
Let us start by recalling the following Hilbert space definitions:
\begin{align}
\mathcal{H}_n &=  {\rm span} \left(\{ u_k[\vartheta]({\bf r}, t)\}\right), \\
\mathcal{H}_{\partial}&={\rm span} \left(\{\partial_\vartheta u_k[\vartheta]({\bf r}, t)\}\right) \setminus \mathcal{H}_n.
\end{align}
We now assume that $m$ is the number of derivatives $\partial_\vartheta u_k[\vartheta]({\bf r}, t)$ that are linearly independent from the $n$ initially populated modes, i.e. ${\rm dim} (\mathcal{H}_{\partial}) = m$.
Accordingly, up to a reordering of the basis $\{\partial_\vartheta u_k[\vartheta]({\bf r}, t)\}$, we can always construct a basis of $\mathcal{H}_\partial$ using the orthonormalized version $u_k^\prime [\vartheta]({\bf r}, t)$ of the derivatives of the first $m$ initially populated modes $\partial_\vartheta u_k[\vartheta]({\bf r}, t)$.

We can now choose the modes $u_k^\prime [\vartheta]({\bf r}, t)$ as the first $m$ among the $N-n$ auxiliary vacuum modes that we use to describe the quantum state of the system in the parameter dependent basis.
In light of this, it is convenient to further decompose the matrix $O_{N-n}$ as
\begin{equation}
O_{N-n} = \begin{pmatrix}
O_\partial & O_{N-n-m}
\end{pmatrix},
\end{equation}
where $O_\partial$ and $O_{N-n-m}$ are matrices of dimensions $2N\times 2m$ and $2N\times 2(N-n-m)$, respectively.
From the orthogonality of $O$, we can obtain the following relations:
\begin{subequations}
\begin{align}
O_{k}^\top  O_{l}  &= \delta_{kl}\mathds{1}_{{\rm dim}(O_k)}, \label{O_properties_1}\\ 
\sum_k O_{k}  O_k^\top &= \mathds{1}_{2N}, \label{O_properties_4}
\end{align}
\label{O_properties}
\end{subequations}
where the sum in Eq.~\eqref{O_properties_4} runs over the total number of column blocks we decomposed the matrix $O$ into.

Let us now compute the derivative of the matrix $O_{n}$. 
In Sec.~\ref{Sec:modes_states}, we have seen that that $O$ is composed by $2 \times 2$ blocks containing the mode overlaps.
Accordingly, it is sufficient to specify the derivative of the $kl$ block of $O_{n}$, which reads
\begin{equation}
\partial_\vartheta \left( O_{n} \right)_{kl} = 
\begin{pmatrix}
\Re \left[(v_l|\partial_\vartheta u_k[\vartheta])\right] & -\Im\left[(v_l|\partial_\vartheta u_k[\vartheta])\right] \\
\Im\left[(v_l|\partial_\vartheta u_k[\vartheta])\right] & \Re \left[(v_l|\partial_\vartheta u_k[\vartheta])\right]
\end{pmatrix}
\label{derivative_On_blocks}
\end{equation}
Combining Eqs.~\eqref{derivative_On_blocks} and \eqref{derivative_modes}, we obtain
\begin{equation}
\partial_\vartheta O_{n} = O_{n} D^\top_n + O_{\partial} D^\top_\partial,
\label{partialO}
\end{equation}
where $D_n$ and  $D_\partial $ are a $2n \times 2n$ and a $2n \times 2m$ matrices, respectively. 
Their $kl$ blocks are given by
\begin{align}
\left(D_n\right)_{kl} = 
\begin{pmatrix}
\Re \left( c_{kl} [\vartheta]\right) & -\Im \left( c_{kl} [\vartheta]\right) \\
\Im \left( c_{kl} [\vartheta]\right) & \Re \left( c_{kl} [\vartheta]\right)
\end{pmatrix}, \label{Cn}\\
\left(D_\partial\right)_{kl} = 
\begin{pmatrix}
\Re \left( c^\prime_{kl} [\vartheta]\right) & -\Im \left( c^\prime_{kl} [\vartheta]\right) \\
\Im \left( c^\prime_{kl} [\vartheta]\right) & \Re \left( c^\prime_{kl} [\vartheta]\right)
\end{pmatrix}. \label{Cpartial}
\end{align}
Using Eqs.~\eqref{partialO} and \eqref{O_properties}, we then obtain
\begin{subequations}
\begin{align}
O^\top_n(\partial_\vartheta O_n)  &= D_n^\top , \\
O^\top_{(N-n)}(\partial_\vartheta O_n)  &= (D_\partial, 0)^\top.
\end{align}
\end{subequations}

Let us conclude this appendix with few words on the coefficients $c_{kl} [\vartheta]$ and $c^\prime_{kl} [\vartheta]$.
The first are simply given by the overlaps of the initially populated modes with their derivatives $c_{kl} [\vartheta] = (u_l[\vartheta]|\partial_\vartheta u_k[\vartheta])$.
On the other hand, there exist several orthonormalization methods that can be used to construct the modes $u_k^\prime [\vartheta]({\bf r}, t)$, leading to different expressions for the coefficients $c^\prime_{kl} [\vartheta]$.
For example, using the Gram-Schmidt procedure, the modes $u_k^\prime [\vartheta]({\bf r}, t)$ can be constructed iteratively as $u_k^\prime [\vartheta]({\bf r}, t) = \tilde{u}_k^\prime [\vartheta]({\bf r}, t)/\sqrt{( \tilde{u}_k^\prime [\vartheta]| \tilde{u}_k^\prime [\vartheta])}$ with
\begin{align}
\tilde{u}_k^\prime [\vartheta]({\bf r}, t) &= \partial_\vartheta u_k [\vartheta]({\bf r}, t) \\
 &\quad - \sum_{j=1}^n ( u_j[\vartheta]|\partial_\vartheta u_k [\vartheta])u_j [\vartheta]({\bf r}, t)\nonumber \\
&\quad - \sum_{j= 1}^{k-1} ( u^\prime_j[\vartheta]|\partial_\vartheta u_k [\vartheta]) u_j^\prime [\vartheta]({\bf r}, t). \nonumber
\end{align}
Accordingly, the coefficients $c^\prime_{kl} [\vartheta]$ are given by
\begin{equation}
c^\prime_{kl} [\vartheta] = 
\begin{cases}
\sqrt{( \tilde{u}_k^\prime [\vartheta]| \tilde{u}_k^\prime [\vartheta])} & {\rm for} \; k = l \\  
( u^\prime_j[\vartheta]|\partial_\vartheta u_k [\vartheta]) & {\rm for} \; k < l \\
0 & k > l
\end{cases},
\end{equation}
resulting in a lower-triangular block matrix $D_\partial$.

\section{Derivative modes for temporal separation estimation}
\label{App:upmd}
\subsection{General expressions}
In this appendix, we construct the orthogonalized first and second derivatives of the modes $u_0[d](t)$ and $u_0[d](t)$.
Let us start by computing the derivatives of Eqs.~\eqref{u0v0} with respect to the parameter $\tau$:
\begin{subequations}
\begin{align}
\partial_d u_0[\tau](t) &= \frac{-\partial_t u(t-\tau/2) + \partial_x u(x+\tau/2) }{2\sqrt{2(1+\delta)}} \nonumber \\ & \quad - \delta^\prime \frac{u_0[\tau](t)}{2(1+\delta)}, \\
\partial_d v_0[\tau](t) &= \frac{-\partial_t u(t-\tau/2) - \partial_t u(t+\tau/2) }{2\sqrt{2(1-\delta)}} \nonumber \\ & \quad + \delta^\prime \frac{v_0[\tau](t)}{2(1-\delta)}, 
\end{align}
\end{subequations}
where we introduced $\delta^\prime = \partial_\tau \delta$.
We assumed that our pulses are symmetric, i.e. $u(t) = u(-t)$. We thus have $(\partial_t u| u) =0$, which, combined with 
\begin{align}
\delta^\prime &= -\int \partial_t u(t-\tau/2) u (t+\tau/2) d t \\
&\; + \int u(t-\tau/2) \partial_t u (t+\tau/2) d t,  \nonumber
\end{align}
implies that $(\partial_\tau v_0| u_0) = (\partial_\tau u_0| v_0) = 0$. 
Consequently, the orthogonalised first derivative modes are simply given by
\begin{subequations}
\begin{align}
u_1[\tau](t) & = \partial_\tau u_0[\tau](t) /\eta_u, \\
v_1[\tau](t) & = \partial_\tau v_0[\tau](t) /\eta_v ,
\end{align}
\end{subequations}
with 
\begin{subequations}
\begin{align} 
\eta_u^2 &= ||\partial_\tau u_0[\tau](t)||^2 = \frac{(\Delta k)^2 - \beta}{4(1+\delta)} - \frac{(\delta^\prime)^2}{4(1+\delta)^2},\\ 
\eta_v^2 &= ||\partial_\tau v_0[\tau](t)||^2 = \frac{(\Delta k)^2 + \beta}{4(1-\delta)} - \frac{(\delta^\prime)^2}{4(1-\delta)^2},
\end{align}
\end{subequations}
where we introduced 
\begin{align}
(\Delta k)^2 &= \int [\partial_t u(t)]^2 dt, \quad {\rm and} \\
\beta &= \int \partial_t u(t-\tau/2)  \partial_t u(t+\tau/2) dt.
\end{align}

We now move to the construction of the orthonormalized second derivatives.
The fact that the modes $u_1[\tau](t)$ and $v_1[\tau](t)$ have, by construction, opposite parity implies that $(\partial_\tau u_1| u_1) = (\partial_\tau v_1| v_1) = 0$, and $(\partial_\tau u_1| v_1) = - (\partial_\tau v_1| u_1)$.
Let us then evaluate $(\partial_d u_1| v_1)$ explicitly 
\begin{align}
(\partial_\tau u_1| v_1) &= \frac{(\partial_\tau u_1| \partial_\tau v_0)}{\eta_v} \\&= \frac{(\partial^2_\tau u_0| \partial_\tau v_0)}{\eta_u \eta_v} - \frac{\eta^\prime_u}{\eta_v \eta_u^2} (\partial_\tau u_0| \partial_\tau v_0)  \nonumber \\&= \frac{(\partial^2_\tau u_0| \partial_\tau v_0)}{\eta_u \eta_v}, \nonumber
\end{align}
where in the last step we used that $u_0[\tau](t)$ and $v_0[\tau](t)$ are even and odd functions of $t$, respectively. 
The second derivative of $u_0(x)$ and $v_0(x)$ can be rewritten as 
\begin{subequations}
\begin{align}
\partial_t^2 u_0[\tau](t) &= \frac{f_u[\tau](t)}{4\sqrt{2(1+\delta)}} -\frac{\delta^\prime}{1+\delta} \partial_d u_0[\tau](t)  \nonumber \\ &\quad+ C_u u_0[\tau](t), \label{d2u0} \\
\partial_t^2 v_0[\tau](t) &= \frac{f_v[\tau](t)}{4\sqrt{2(1-\delta)}} +\frac{\delta^\prime}{1-\delta} \partial_d v_0[\tau](t) \nonumber \\ & \quad+ C_v v_0[\tau](t), \label{d2v0}
\end{align}
\end{subequations}
with 
\begin{align}
f_u[\tau] (t) &= \partial^2_t u(t-\tau/2) + \partial^2_t u(t+\tau/2), \\
f_v[\tau] (t) &= \partial^2_t u(t-\tau/2) - \partial^2_t u(t+\tau/2),
\end{align}
and
\begin{align}
C_u &= \frac{(\delta^\prime)^2 - 2(1+\delta)\delta^{\prime\prime}}{4(1+\delta)^2},\\
C_v& = \frac{(\delta^\prime)^2 + 2(1- \delta)\delta^{\prime\prime}}{4(1-\delta)^2},
\end{align}
where we have introduced the second derivative of the overlap parameter $\delta^{\prime \prime} = \partial_\tau^2 \delta$.
Using $(\partial_\tau u_0|\partial_\tau v_0) = (u_0|\partial_\tau v_0) =0$, from Eq.~\eqref{d2u0} we have 
\begin{align}
&(\partial_d u_1| v_1) \propto \\
%&\int \partial^2_t\left( u(t-\tau/2) + u(t+\tau/2)\right)\partial_t\left( u(t-\tau/2) + u(t+\tau/2)\right) dt \nonumber \\
&\int \partial_t^2 u(t+\tau) \partial_t u(t) dt + \int \partial_t^2 u(-t-\tau) \partial_t u(-t) dt \nonumber \\
&= \int \partial_t^2 u(t+\tau) \partial_t u(t) dt \nonumber - \int \partial_t^2 u(t+\tau) \partial_t u(t) dt\\& = 0, \nonumber
\end{align}
where we used that the pulse shape $u(t)$ is an even function of $t$.
Therefore, the orthogonalized second derivative modes are given by
\begin{align}
u_2[\tau](t) & = (\partial_d u_1[\tau](t) - \xi_u u_0([\tau](t) )/\zeta_u, \\
v_2[\tau](t)  & = (\partial_d v_1[\tau](t)  - \xi_v v_0[\tau](t) )/\zeta_v, 
\end{align}
with $\xi_u = (\partial_\tau u_1| u_0)$, $\xi_v = (\partial_\tau v_1| v_0)$,  $\zeta_u = ||\partial_\tau u_1[\tau(t) - \xi_u u_0[\tau](t)||$ and $\zeta_v = ||\partial_\vartheta v_1[\vartheta](t) - \xi_v v_0[\vartheta](t)||$.

Let us explicitly calculate $\xi_u$ and $\xi_v$. 
This can be achieved by using Eqs.~\eqref{d2u0}, \eqref{d2v0} and by noting that
\begin{align}
\int &\partial_t^2\left[ u(t-\tau/2) \pm u(t+\tau/2)\right] \\ \quad &\times\left[ u(t-\tau/2) \pm u(t-\tau/2)\right] dt \nonumber\\
&= -\int  \left[\partial_tu(t-\tau/2) \pm  \partial_tu(t+\tau/2)\right]^2 dt \nonumber\\
&=-2(\Delta k)^2 \mp 2\beta, \nonumber
\end{align}
where we used partial integration and made the reasonable assumption that the pulse shape $u(t)$ goes to zero at infinity. 
We then get 
\begin{subequations}
\begin{align}
\xi_u &= \frac{1}{\eta_u}\left(\frac{(\delta^\prime)^2 - 2(1+\delta)\delta^{\prime \prime}}{4(1+\delta)^2} - \frac{(\Delta k)^2 +\beta}{4(1+\delta)}\right) \\ & =\frac{(\delta^\prime)^2 - 2(1+\delta)\delta^{\prime \prime} -(1+\delta)\left((\Delta k)^2+\beta\right)}{(1+\delta)\sqrt{(1+\delta)\left((\Delta k)^2-\beta\right)-(\delta^\prime)^2}}, \nonumber\\
\xi_v &= \frac{1}{\eta_v}\left(\frac{(\delta^\prime)^2 + 2(1-\delta)\delta^{\prime \prime}}{4(1-\delta)^2} - \frac{(\Delta k)^2 - \beta}{4(1-\delta)}\right)  \\ & = \frac{(\delta^\prime)^2 + 2(1-\delta)\delta^{\prime \prime} -(1+\delta)\left((\Delta k)^2-\beta\right)}{(1-\delta)\sqrt{(1-\delta)\left((\Delta k)^2+\beta\right)-(\delta^\prime)^2}}. \nonumber
\end{align}
\end{subequations}

Let us now compute explicitly the normalization constants of the modes $u_2[\tau](t)$ and $v_2[\tau](t)$
\begin{subequations}
\begin{align}
\zeta_u^2 &= || \partial_\tau u_1[\tau](t) - \xi_u u_0[\tau](t)||^2 \\
&= ||\partial_\tau u_1[\tau](t)||^2 +\xi_u^2 - 2\xi_u (\partial_\tau u_1| u_0) \nonumber \\
& = ||\partial_\tau u_1[\tau](t)||^2 -\xi_u^2, \nonumber \\ 
\zeta_v^2 &= || \partial_\tau v_1[\tau](t) - \xi_v v_0[\tau](t)||^2 \\
&= ||\partial_\tau v_1[\tau](t)||^2 +\xi_v^2 - 2\xi_v (\partial_\tau v_1| v_0)\nonumber \\
&= ||\partial_\tau v_1[\tau](t)||^2 -\xi_v^2. \nonumber 
\end{align}
\end{subequations}
We can then expand
\begin{subequations}
\begin{align}
||\partial_\tau u_1[\tau](t)||^2 &= \left|\left| \frac{\partial^2_\tau u_0[\tau](t)}{\eta_u} - \frac{(\partial_\tau \eta_u)^2}{\eta_u^2}\partial_\tau u_0 [\tau](t) \right|\right|^2 \\ 
&= \frac{||\partial^2_\tau u_0||^2+ (\partial_\tau \eta_u)^2}{\eta_u^2} \nonumber \\ 
&\;- 2\frac{\partial_\tau \eta_u}{\eta_u^3}(\partial_\tau^2 u_0| \partial_\tau u_0), \nonumber \\
||\partial_\tau v_1[\tau](t)||^2 &= \left|\left| \frac{\partial^2_\tau v_0[\tau](t)}{\eta_v} - \frac{(\partial_\tau \eta_v)^2}{\eta_v^2}\partial_\tau v_0[\tau](t) \right|\right|^2 \\&= \frac{||\partial^2_\tau v_0 [\tau](t)||^2+ (\partial_\tau \eta_v)^2}{\eta_v^2} \nonumber \\ 
&\; - 2\frac{\partial_\tau \eta_v}{\eta_v^3}(\partial_\tau^2 v_0| \partial_\tau v_0). \nonumber
\end{align}
\end{subequations}
From Eqs.~\eqref{d2u0} and \eqref{d2v0}, we then have 
\begin{subequations}
\begin{align}
||\partial^2_\tau u_0[\tau](t)||^2 &= \frac{||f_u[\tau](t)||^2}{32(1+\delta)} +\frac{(\delta^\prime)^2\eta_u^2}{(1+\delta^2)} + C_u^2 \\
&\; -\frac{\delta^\prime (f_u|\partial_\tau u_0)}{4\sqrt{2}(1+\delta)^{3/2}} - C_u\frac{2(\Delta k)^2 +2\beta}{\sqrt{2(1+\delta)}} \nonumber \\ 
||\partial^2_\tau v_0[\tau](t)||^2 &=\frac{||f_v[\tau](t)||^2}{32(1-\delta)}+\frac{(\delta^\prime)^2\eta_v^2}{(1-\delta^2)} + C_v^2 \\ &\;+ \frac{\delta^\prime (f_v|\partial_\tau v_0)}{4\sqrt{2}(1-\delta)^{3/2}} -C_v\frac{2(\Delta k)^2 -2\beta}{\sqrt{2(1-\delta)}}, \nonumber
\end{align}
\end{subequations}
with
\begin{subequations}
\begin{align}
||f_u[\tau](t)||^2 &= 2(\sigma +\epsilon) \\
||f_v[\tau](t)||^2 &= 2(\sigma -\epsilon),
\end{align}
\end{subequations}
and 
\begin{subequations}
\begin{align}
(f_u|\partial_du_0) & = \frac{(\Delta k)^2 +\beta}{1+\delta}\delta^\prime +\frac{\beta^\prime}{\sqrt{2(1+\delta)}},\\
(f_v|\partial_dv_0)&=\frac{(\Delta k)^2 -\beta}{1-\delta}\delta^\prime+\frac{\beta^\prime}{\sqrt{2(1-\delta)}},
\end{align}
\end{subequations}
where we defined 
\begin{subequations}
\begin{align}
\sigma &= \int \left|\partial_t^2 u(t)\right|^2 dt,\\
\epsilon &= \int \partial_t^2 u(t-\tau/2) \partial_t^2 u(t-\tau/2) d x.
\end{align}
\end{subequations}

Since the expressions of the mode quantities (especially $\zeta_u$ and $\zeta_v$) computed above  for a generic pulse shape $u(t)$ are fairly complicated, we present their explicit expressions for a Gaussian pulse $u(t) = e^{-t^2/2 w^2}/(\pi w^2)^{1/4}$ in the following:
\begin{align}
\eta_u  &= \frac{\sqrt{\tau ^2+4 w^2 \sinh \left(\frac{\tau ^2}{4 w^2}\right)} \text{sech}\left(\frac{\tau ^2}{8 w^2}\right)}{8 w^2},\\
\eta_v &= \frac{\sqrt{4 w^2 \sinh \left(\frac{\tau ^2}{4 w^2}\right)-\tau ^2} \text{csch}\left(\frac{\tau ^2}{8 w^2}\right)}{8 w^2},\\
\xi_u &= -\frac{\sqrt{\tau ^2+4 w^2 \sinh \left(\frac{\tau ^2}{4 w^2}\right)} \text{sech}\left(\frac{\tau ^2}{8 w^2}\right)}{8 w^2},\\
\xi_v &=  -\frac{\sqrt{4 w^2 \sinh \left(\frac{\tau ^2}{4 w^2}\right)-\tau ^2} \text{csch}\left(\frac{\tau ^2}{8 w^2}\right)}{8 w^2},\\ 
\zeta^2_u &= \frac{2 w^2 \sinh \left(\frac{\tau ^2}{2 w^2}\right)-\tau ^2 \left(1+\cosh \left(\frac{\tau ^2}{4 w^2}\right)\right)}{\left(\tau ^2+4 w^2 \sinh \left(\frac{\tau ^2}{4 w^2}\right)\right)^2}\\ 
 &\quad +\frac{\left(\tau^4+16 w^4\right) \sinh \left(\frac{\tau ^2}{4 w^2}\right)}{4 w^2 \left(\tau ^2+4 w^2 \sinh \left(\frac{\tau ^2}{4 w^2}\right)\right)^2} \nonumber\\
\zeta_v^2 &= \frac{2 w^2 \sinh \left(\frac{\tau ^2}{2 w^2}\right)-\tau ^2 \left(1-\cosh \left(\frac{\tau ^2}{4 w^2}\right)\right)}{\left(\tau ^2-4 w^2 \sinh \left(\frac{\tau ^2}{4 w^2}\right)\right)^2}\\ 
 &\quad -\frac{\left(\tau^4+16 w^4\right) \sinh \left(\frac{\tau ^2}{4 w^2}\right)}{4 w^2 \left(\tau ^2-4 w^2 \sinh \left(\frac{\tau ^2}{4 w^2}\right)\right)^2} \nonumber.
\end{align}
Note that the fact that $\eta_{u,v}^2 = \xi_{u,v}^2$ is a peculiarity of Gaussian pulses and it is not true in general.

\subsection{Small $\tau$ behaviour}
Arguably, the most interesting regime for temporal separation estimation is that of small $\tau$. Therefore, in the following, we discuss the behaviour of the quantities computed above for $\tau \to 0$.
Let us start by considering the following series expansions
\begin{subequations}
\begin{align}
\delta &= 1 - (\Delta k)^2\frac{\tau^2}{2} + \sigma \frac{\tau^4}{24} + \mathcal{O}(\tau^6), \\
\beta &= (\Delta k)^2 -\sigma \frac{\tau^2}{2} + \mathcal{O}(\tau^4),\\
\epsilon &= \sigma + \mathcal{O}(\tau^2).
\end{align}
\label{small_tau}
\end{subequations}
Using Eqs.~\eqref{small_tau}, it is possible to show that for $\tau \sim 0$ we have $\eta_u \sim \eta_v \sim \xi_u \sim \xi_v \sim \zeta_u \sim \zeta_v \sim \tau$.
For example, for Gaussian pulses we have
\begin{subequations}
\begin{align}
    \eta_u  &= \frac{\tau }{4 \sqrt{2} w^2}+\mathcal{O}\left(\tau ^2\right), \\
    \eta_v  &= \frac{\tau }{4 \sqrt{6} w^2}+\mathcal{O}\left(\tau ^2\right), \\
    \xi_u   &= -\frac{\tau }{4 \sqrt{2} w^2}+\mathcal{O}\left(\tau ^2\right), \\
    \xi_v   &= -\frac{\tau }{4 \sqrt{6} w^2}+\mathcal{O}\left(\tau ^2\right), \\
    \zeta_u &= \frac{\tau }{4 \sqrt{3} w^2}+\mathcal{O}\left(\tau ^2\right), \\
    \zeta_v &= \frac{\tau }{4 \sqrt{5} w^2}+\mathcal{O}\left(\tau ^2\right).
\end{align}
\end{subequations}
This behaviour implies that the contribution to the QFI coming from the coefficients $b_l^{jk}$ and $d_l^{jk}$ vanishes for $\tau \to 0$ (see Eqs.\eqref{abcd} and App.~\ref{App:QFI_tau_th_squ}).

\section{Calculation of the QFI~\eqref{QFI_tau_therm_squeez}}
\label{App:QFI_tau_th_squ}
In this Appendix, we explicitly compute the coefficients~\eqref{abcd} that lead to the QFI~\eqref{QFI_tau_therm_squeez} for the estimation of the temporal separation $\tau$ between two incoherent thermal pulses aided by two squeezed vacuum states in the derivative modes $u_1[\tau](t)$ and $v_1[\tau](t)$ defined by the covariance matrices $V_0$~\eqref{V_0_tau} and $V_1~\eqref{V_1_tau}$. 

First, we note that $V_0$~\eqref{V_0_tau} is already in Williamson form. Therefore, the symplectic matrix $S_0$ entering in Eqs.~\eqref{abcd} is the identity $S_0 = \mathds{1}_4$, while for the symplectic eigenvalues we have $\nu_0^0 = 2N_0(1+\delta) + 1$ and  $\nu_0^1 = 2N_0(1-\delta) + 1$. 
On the other hand, the Williamson decomposition $V_1$~\eqref{V_1_tau} is achieved by the squeezing matrix 
\begin{equation}
    S_1 = \begin{pmatrix}
    e^{-r} & 0 & 0 & 0 \\
    0 & e^{r}& 0 & 0 \\
    0 & 0 & e^{-r} & 0\\
    0 & 0 & 0 &  e^{r} \\
    \end{pmatrix},
    \label{S_1_tau}
\end{equation}
with symplectic eigenvalues $\nu_1^{0,1} = 1$.

To compute the coefficients $a_l^{jk}$, we need the derivative of the matrix $V_0$~\eqref{V_0_tau}. 
The latter depends on the temporal separation $\tau$ only through the overlap parameter $\delta$. Accordingly, we have
\begin{equation}
    \partial_\tau V_0 =
    \begin{pmatrix}
    2 N_0 (\partial_\tau \delta) \mathds{1}_2 & 0 \\
    0 & -2 N_0 (\partial_\tau \delta) \mathds{1}_2
    \end{pmatrix}.
\end{equation}
Since $\partial_\tau V_0$ is diagonal, the only nonzero $a_l^{jk}$ coefficients are
\begin{equation}
    \left(a^{00}_2\right)^2 = \left(a^{11}_2 \right)^2 = 8 N_0 (\partial_\tau \delta).
    \label{a_tau}
\end{equation}

Using $S_0 = \mathds{1}_4$, and the fact that all matrices $D_\xi$, $D_\eta$, and $V_0$ are diagonal, we have $
\left(b_l^{jk} \right)^2 = 2 \left( \tr \left\{  A_l^{(jk)}M_b\right\} \right)^2,
$
where we introduced the matrix
\begin{equation}
    M_b = \begin{pmatrix}
    X_u + Y_u & 0 & 0 & 0\\
    0 & X_u - Y_u & 0 & 0\\
    0 & 0 &  X_v + Y_v & 0\\
    0 & 0 & 0 & X_v - Y_v
    \end{pmatrix},
\end{equation}
with 
\begin{subequations}
\begin{align}
    X_u & = N_0(1+\delta)\eta_u \cosh r, \\
    X_v & = N_0(1-\delta)\eta_v \cosh r, \\
    Y_u & = \left[N_0(1+\delta)\eta_u -\xi_u \right]\sinh r,\\
    Y_v & = \left[N_0(1-\delta)\eta_v -\xi_v \right]\sinh r.
\end{align}
\end{subequations}
Consequently, the only nonzero $b_l^{jk}$ coefficients are
\begin{subequations}
\begin{align}
\left(b_1^{00}\right)^2 & = 8\left[N_0(1+\delta)\eta_u -\xi_u\right]^2 \sinh^2 r,\\
\left(b_2^{00}\right)^2 & = 8N_0^2(1+\delta)^2\eta_u^2 \cosh^2 r, \\
\left(b_1^{11}\right)^2 & = 8\left[N_0(1-\delta)\eta_v -\xi_v\right]^2 \sinh^2 r, \\ 
\left(b_2^{11}\right)^2 & = 8N_0^2(1-\delta)^2\eta_v^2 \cosh^2 r.
\end{align}
\label{b_tau}
\end{subequations}

We assumed the covariance matrix $V_1$ of the derivative modes $u_1[\tau](t)$ and $v_1[\tau](t)$ to be parameter independent, i.e. $\partial_\tau V_1 = 0$, which leads to $c_l^{(jk)} = 0$ for all $l, j$ and $k$.

Finally, we can write the $d_l^{jk}$ coefficients as 
\begin{equation}
    \left(d_l^{jk} \right)^2 = 2 \left( \tr \left\{  A_l^{(jk)}M_d\right\} \right)^2
\end{equation}
with 
\begin{equation}
M_d = D_\zeta (V_1 - \mathds{1}_4) S_1^{-1} = 2 \sinh r\begin{pmatrix}
\zeta_u \sigma_z & 0 \\ 
0 & \zeta_v \sigma_z
\end{pmatrix}.
\end{equation}
As consequence, the only nonzero $d_l^{jk}$ coefficients are
\begin{subequations}
\begin{align}
\left(d_1^{00}\right)^2  & = 8 \zeta_u^2 \sinh^2 r,\\
\left(d_1^{11}\right)^2  & = 8 \zeta_v^2 \sinh^2 r.
\end{align}
\label{d_tau}
\end{subequations}
Substituting the coefficients in Eqs.~\eqref{a_tau}, \eqref{b_tau} and \eqref{d_tau} into Eq.~\eqref{F_v_uv}, we then obtain the QFI~\eqref{QFI_tau_therm_squeez}.

\bibliography{SR_long}{}
\bibliographystyle{apsrev4-2}
\end{document}